\documentclass[12pt, draftclsnofoot, onecolumn]{IEEEtran}
\UseRawInputEncoding
\normalsize
\usepackage{gensymb}
\usepackage{lineno}
\usepackage{fancyhdr}
\usepackage{amsmath, amsthm, amssymb}
\usepackage{amsfonts}
\usepackage[dvips]{graphicx}
\usepackage[normalem]{ulem}
\usepackage{balance}
\usepackage{epsfig}
\usepackage[lined,boxed,commentsnumbered]{algorithm2e}
\usepackage{color}
\usepackage{caption}
\usepackage{enumerate}
\usepackage{latexsym}
\usepackage{graphics}
\usepackage{graphicx}
\usepackage{subfigure}
\usepackage{cite}
\usepackage{mathrsfs}
\usepackage{algorithmic}
\usepackage{stfloats}
\usepackage{extarrows}
\usepackage{upgreek}
\usepackage{lineno}
\usepackage{caption}
\usepackage{booktabs}

\usepackage{amsmath}
\usepackage{setspace}
\usepackage{array}
\usepackage{soul} % 导入 soul 包
\usepackage{color, xcolor}

\usepackage{graphicx} %use graph format
\usepackage{epstopdf}
\usepackage{float}
\usepackage{subfigure}
\usepackage{yhmath}
\usepackage{fancyhdr}

\newcommand{\vect}[1]{\mathbf{#1}}

\begin{document}
\captionsetup[figure]{labelfont={small},textfont={small},labelformat={default},labelsep=period,name={Fig.},justification={raggedright}}

\title{Joint Optimization of Resource Allocation and Trajectory Control for Mobile Group Users in Fixed-Wing UAV-Enabled Wireless Network}
\author{
Xuezhen~Yan,
Xuming~Fang, {\em Senior Member, IEEE},
Cailian~Deng,
Xianbin Wang, {\em Fellow, IEEE}

\thanks{%The work of X. Yan, X. Fang, and C. Deng was supported in part by NSFC under Grant 62071393, NSFC High-Speed Rail Joint Foundation under Grant U1834210, Sichuan Provincial Applied Basic Research Project under Grant 2020YJ0218, and Fundamental Research Funds for the Central Universities under Grant 2682021CF019. \textit{(Corresponding author: Xuming Fang.)}

X. Yan,  X. Fang, and C. Deng are with the Key Laboratory of Information Coding and Transmission, Southwest Jiaotong University, Chengdu 611756, China. E-mail: xuezhenyan@my.swjtu.edu.cn; xmfang@swjtu.edu.cn; dengcailian@my.swjtu.edu.cn.

X. Wang is with the Department of Electrical and Computer Engineering, Western University, London, ON N6A 5B9, Canada. E-mail: xianbin.wang@uwo.ca.

}

}
\maketitle
\thispagestyle{fancy}

\lfoot{~\copyright~2022 IEEE. This work has been submitted to the IEEE for possible publication. Copyright may be transferred without notice, after which this version may no longer be accessible.}

\cfoot{}

\renewcommand{\headrulewidth}{0mm}
\begin{abstract}
Owing to the controlling flexibility and cost-effectiveness, fixed-wing unmanned aerial vehicles (UAVs) are expected to serve as flying base stations (BSs) in the air-ground integrated network. By exploiting the mobility of UAVs, controllable coverage can be provided for mobile group users (MGUs) under challenging scenarios or even somewhere without communication infrastructure. However, in such dual mobility scenario where the UAV and MGUs are all moving, both the non-hovering feature of the fixed-wing UAV and the movement of MGUs will exacerbate the dynamic changes of user scheduling, which eventually leads to the degradation of MGUs' quality-of-service (QoS). In this paper, we propose a fixed-wing UAV-enabled wireless network architecture to provide moving coverage for MGUs. In order to achieve fairness among MGUs, we maximize the minimum average throughput between all users by jointly optimizing the user scheduling, resource allocation, and UAV trajectory control under the constraints on users' QoS requirements, communication resources, and UAV trajectory switching. Considering the optimization problem is mixed-integer non-convex, we decompose it into three optimization subproblems. An efficient algorithm is proposed to solve these three  subproblems alternately till the convergence is realized. Simulation results demonstrate that the proposed algorithm can significantly improve the minimum average throughput of MGUs.
\end{abstract}


\section{Introduction} \label{sec-intro}

%\IEEEPARstart
With the rapid development of wireless communications, the unmanned aerial vehicle (UAV)-enabled network is expected to become an essential component of the future sixth-generation (6G) wireless networks to achieve ubiquitous connectivity \cite{6GGiordani,ZengMag,BeidaLOS}. Compared with terrestrial wireless communication systems, one major advantage of UAV base stations (BSs) is their high probability of providing line-of-sight (LOS) communication links to ground users, which directly alleviates the challenge of extremely weak signal strength at the receivers caused by shadowing and fading in urban or mountainous areas \cite{BeidaLOS}. Besides, due to the strong environmental adaptability and controllable mobility, utilizing UAV BSs to provide moving coverage for mobile users can be widely applied in emergency rescue \cite{AdaptiveCoverage,Rendezvous}, transportation \cite{MozaffariST}, and surveillance scenarios \cite{WuQingMul}, etc.

In general, UAVs used in wireless networks can be divided into rotary-wing and fixed-wing categories based on the aerodynamic and architectural differences. The choice of UAVs as aerial platforms for BSs mainly depends on the application needs and the cost constraint. Rotary-wing UAVs are mostly used where they are expected to move according to the predetermined trajectory or hover steadily in a fixed position \cite{ZengRotary}. Unfortunately, the flight endurance time of rotary-wing UAVs is only dozens of minutes due to their limited battery capacity. Therefore, the utilization of fixed-wing UAVs has attracted increasing attention in 6G wireless networks \cite{LiuJunyu,CurvatureLet,BeiyouForwarding,ZengYongEE}. Thanks to the greater payload and longer endurance, fixed-wing UAVs are more promising in serving as aerial BSs compared with rotary-wing UAVs.

Recently, trajectory optimization of fixed-wing UAVs has been actively investigated in UAV-enabled wireless networks. Specifically, UAV trajectory optimizations with fixed altitude in a two-dimensional (2D) plane have been extensively explored in \cite{ZengYongEE,BeidaRelay,DubinsAut,DubinsICUAS,SensorFixCur}, while three-dimension (3D) UAV trajectory optimizations have been investigated in \cite{FixSwarms,3DTWC,DubinsEmergingTec}. From the perspective of aerodynamics \cite{LiuJunyu}, trajectory planning of the fixed-wing UAV is mainly limited by non-hovering features and bounded flight curvature. In order to better leverage the maneuverability of UAVs, the path elongation problem for Dubins vehicles with maximum curvature in a 2D plane was studied in \cite{DubinsAut}, providing ideal reference trajectories and expected lengths for Dubins vehicles to accurately control the arrival time. In \cite{DubinsICUAS}, a path generator suitable for fixed-wing UAVs was proposed by combining the Dubins path generation method with Lyapunov-based nonlinear path tracking control. Due to the unique kinematic characteristics of fixed-wing UAVs, such as minimum turning radius, a locally adjustable continuous curvature bounded path planning algorithm was introduced in \cite{SensorFixCur}. In \cite{DubinsEmergingTec}, a UAV trajectory control scheme based on the 3D Dubins curve to smooth the trajectory and constrained by the UAV kinetic properties was investigated for complex urban environments. However, the aforementioned studies on UAV trajectory optimization only focus on the maneuverability of fixed-wing UAVs. The limitations of communication requirements are often ignored during the UAV trajectory planning, which may lead to the degradation of network quality-of-service (QoS).

Since the onboard communication resources (i.e., the total bandwidth and transmit power) of the UAV are very limited, it is crucial to allocate resources effectively to improve communication capabilities in multi-user networks. Therefore, research topics on communication resource allocation in UAV-enabled wireless networks, e.g., maximizing the minimum user throughput and  minimizing the service latency, have been extensively studied \cite{Wuqing,LiRuide,DengCailian}. In \cite{Wuqing}, the UAV was dispatched as a mobile BS to serve ground users with service delay constraints, and the minimum average throughput of all users was maximized by jointly optimizing the orthogonal frequency-division multiple access (OFDMA) resource allocation and UAV trajectory. To achieve fairness in secure communication, the communication/jamming subcarrier allocation strategy and UAV trajectory are jointly optimized to maximize the average minimum secrecy rate of each user \cite{LiRuide}. Such optimization methods can ensure fairness in multi-user networks by redistributing resources from strong users to weak users. In \cite{DengCailian}, with the constraints of communication and computation resources, delay, and energy consumption, a joint optimization algorithm for deep neural networks model decision, resource allocation, and UAV trajectory design was proposed to minimize the service latency.

Despite many studies on UAV mobility and resource allocation in static user scenarios mentioned above, channel conditions of wireless communication systems can also be affected by user mobility, leading to dramatic fluctuation of communication channel conditions. There are only few recent researches on user mobility prediction in air-ground wireless networks. The impact of terrestrial user mobility, propagation environment, and channel fading on air-to-ground communication outage performance were analyzed in \cite{Outage,3DMobUAV,MobPreCN}. To ensure timely delivery of data  in mobile ad hoc networks where UAVs serve as message ferries, an opportunistic data delivery scheme based on mobility prediction of users and UAV placement design was proposed in \cite{Rendezvous}. In addition, in realistic network scenarios, such as fleet transportation and disaster relief operations, the mobile users are often involved in group activities and exhibit common mobility behavior. A novel routing scheme was proposed in \cite{RGPMTWC}, which can obtain an efficient throughput-delay tradeoff for mobile group users (MGUs) in mobile ad hoc networks with reference point group mobility (RPGM). To predict network partitions and reduce the number of interrupts, a simple and effective data clustering algorithm was designed in \cite{GroupmobICC}, which can accurately determine the mobile groups and estimate the characteristic parameters of each group under the condition of given mobile users' speed in wireless ad hoc networks. However, these studies on group mobility mainly focus on the communication connections between users in mobile ad hoc networks. How to improve the communication performance in an air-ground system with dual mobility of the fixed-wing UAV and ground users has become an urgent problem to be solved.

Motivated by the above, we explore the fixed-wing UAV wireless network for MGUs in this paper, where the UAV is launched as a mobile BS. The link quality between the UAV and ground MGUs varies with the flying trajectory, coverage range, and user mobility. To adapt to the variations and achieve fairness between MGUs, we jointly optimize the user scheduling, resource allocation, and UAV trajectory to maximize the minimum average throughput among users in the UAV-enabled wireless network. The main contributions of this paper are summarized as follows:

\begin{itemize}
  \item Considering realistic application scenarios, we propose a novel fixed-wing UAV-enabled wireless network architecture, where the UAV is equipped with BS to provide moving coverage for ground MGUs. Moreover, we combine the communication resources, users' QoS requirements, and UAV trajectory switching constraints to maximize the minimum average throughput among users, which has been rarely investigated in UAV-enabled networks.
  \item Since the optimization problem is mixed-integer non-convex, we decompose it into three more tractable subproblems. An iterative optimization algorithm is further proposed to provide a feasible solution. Firstly, we obtained the user scheduling by using variable relaxation. Secondly, Lagrange duality is applied to jointly optimize bandwidth allocation and power control. Thirdly, the UAV trajectory control subproblem is solved by successive convex approximation (SCA). Finally, these subproblems are alternatively iterated till the convergence is achieved.
  \item Simulation results verify that our proposed iterative algorithm has good convergence performance. Through trajectory optimization, the fixed-wing UAV switches trajectory conforming to the movement characteristics of ground MGUs to provide better channel conditions and obtain higher user throughput gain. Moreover, the user scheduling and resource allocation are optimized according to the wireless channel conditions and the available resources of the UAV BS, greatly enhancing the system performance.
\end{itemize}

The remainder of this paper is organized as follows. In Section II, we introduce the system model and formulate the optimization problem. In Section III, we propose an efficient iterative optimization algorithm. In Section IV, we prove that the proposed algorithm has good convergence and effectiveness through simulation results. Finally, we summarize the work and propose some future work prospects in Section V.

%\begin{figure}[!htbp]
%\renewcommand{\figurename}{Fig.}
%  \centering
%  \includegraphics[width=3.5in]{Fig1.eps}
%  \caption{A Fixed-wing UAV-enabled mobile wireless network.}
%\end{figure}
%
%\begin{figure}[h]
%\centering
%\includegraphics[width=1\columnwidth]{Fig2.eps}
%\caption{Fixed-wing UAV trajectory switching (planform).}
%\label{Fig2}
%\end{figure}

\section{System Model and Problem Formulation}\label{section:System Model}

\subsection{System Model}

As shown in Fig. 1 (a), a fixed-wing UAV is employed as an aerial BS to cover MGUs in the downlink scenario. We assume that MGUs move in the area according to the RPGM model. To represent the group mobility of MGUs, the model defines a logical reference center whose movement is followed by all users in the group \cite{GroupmobICC}. In this paper, we set the initial distribution center of MGUs, namely User Center (UC), as the reference center. Furthermore, it is assumed that the UAV can obtain the location of MGUs through its wireless sensing capability or from the positioning information of global positioning system (GPS) delivered by MGUs. In this way, the UAV can estimate the movement trajectory of MGUs over a period of time through mobility prediction based on the historical data \cite{MobPreCN} and then set its own flying trajectory. Since the fixed-wing UAV cannot hover and requires larger centripetal force to maintain a more curved trajectory, too small turning radius may cause the roll angle to exceed safety limits \cite{CurvatureLet}. Here, we adopt the Dubins switch trajectory model (DSTM) \cite{DubinsICUAS} for the UAV to provide moving coverage for MGUs. In general, the Dubins path consists of different path segments, i.e., arcs/circles and lines. We take the right-straight-right (RSR) model \cite{DubinsICUAS} as an example for analysis, which is a typical Dubins path generated by clockwise rotation. When the UAV flies in the arc/circle path segment, the UAV takes UC as its flying center on the 2D top view projection plane. The flying radius of the UAV in the arc/circle path segments depends on the distribution of MGUs and the threshold of safe turning radius. For simplicity, we assume that the fixed-wing UAV suspends communication with MGUs when switching trajectory, and the time of switching trajectory along a straight line can be ignored \cite{DubinsAut}.
%\begin{figure}[!t]
%\centering
%\begin{minipage}[t]{0.48\textwidth}
%\centering
%\includegraphics[width=6cm]{Fig1.eps}
%\caption{A fixed-wing UAV-enabled downlink mobile wireless network.}
%\end{minipage}
%\begin{minipage}[t]{0.48\textwidth}
%\centering
%\includegraphics[width=6cm]{Fig2.eps}
%\caption{The RSR model for the fixed-wing UAV (planform).}
%\end{minipage}
%\end{figure}

%\begin{figure}[!t]
%\centering
%\subfigure[A fixed-wing UAV-enabled downlink mobile wireless network.]{
%\label{Fig1.1}
%\begin{minipage}[t]{0.45\textwidth}
%\centering
%\includegraphics[width=1in]{Fig1.eps}
%\end{minipage}
%}
%\subfigure[The RSR model for the fixed-wing UAV (planform).]{
%\label{Fig1.2}
%\begin{minipage}[t]{0.45\textwidth}
%\centering
%\includegraphics[width=1in]{Fig2.eps}
%\end{minipage}
%}
%\centering
%\caption{ 3D and 2D illustration of the fixed-wing UAV-enabled downlink mobile wireless network.}
%\label{Fig5}
%\end{figure}
\begin{figure}[!t]
\renewcommand{\figurename}{Fig.}
\centering
\subfigure[A fixed-wing UAV-enabled downlink mobile wireless network.]{
\label{Fig1.1}
\includegraphics[width=0.45\textwidth]{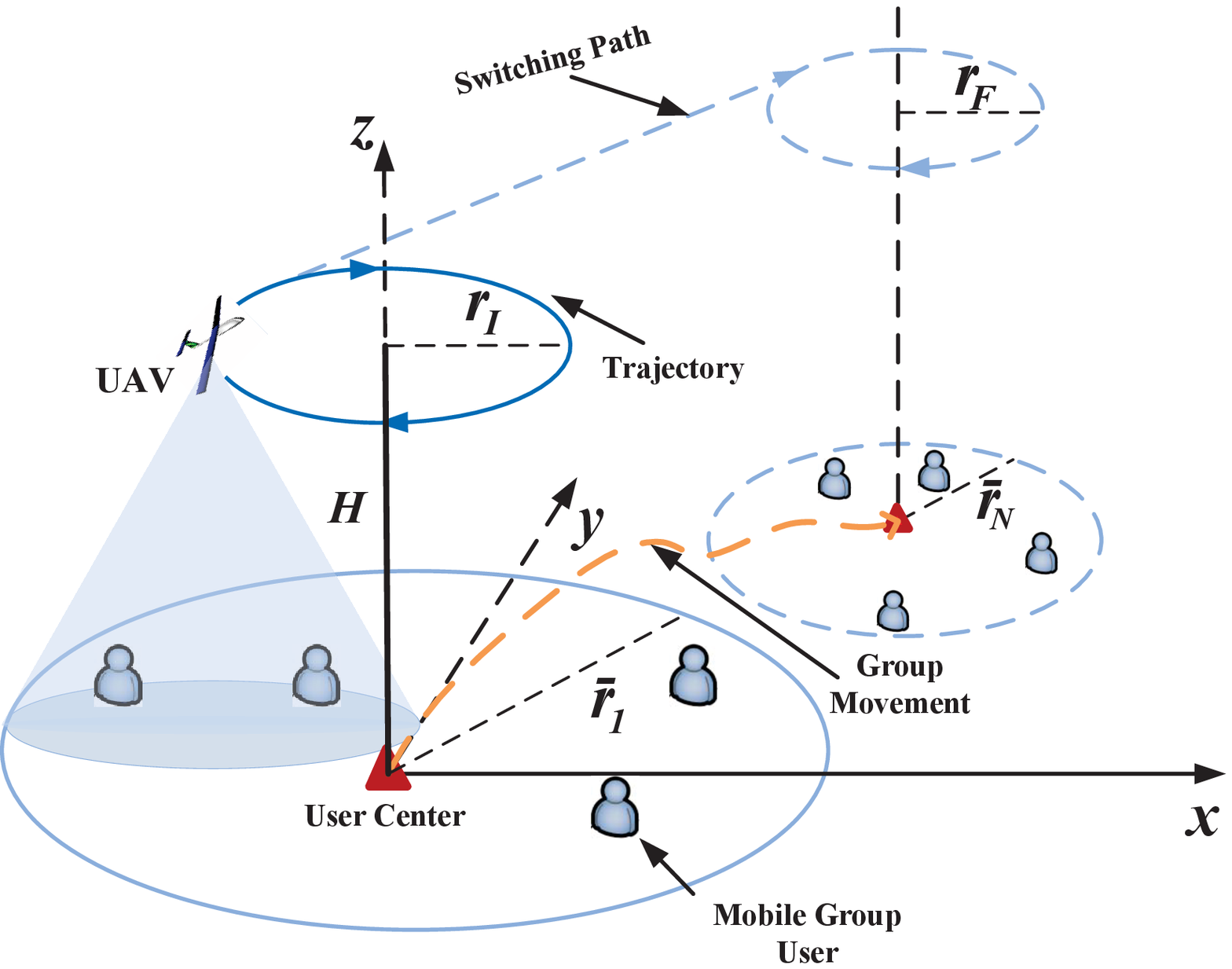}}
\subfigure[The RSR trajectory model for the fixed-wing UAV (planform).]{
\label{Fig1.2}
\includegraphics[width=0.45\textwidth]{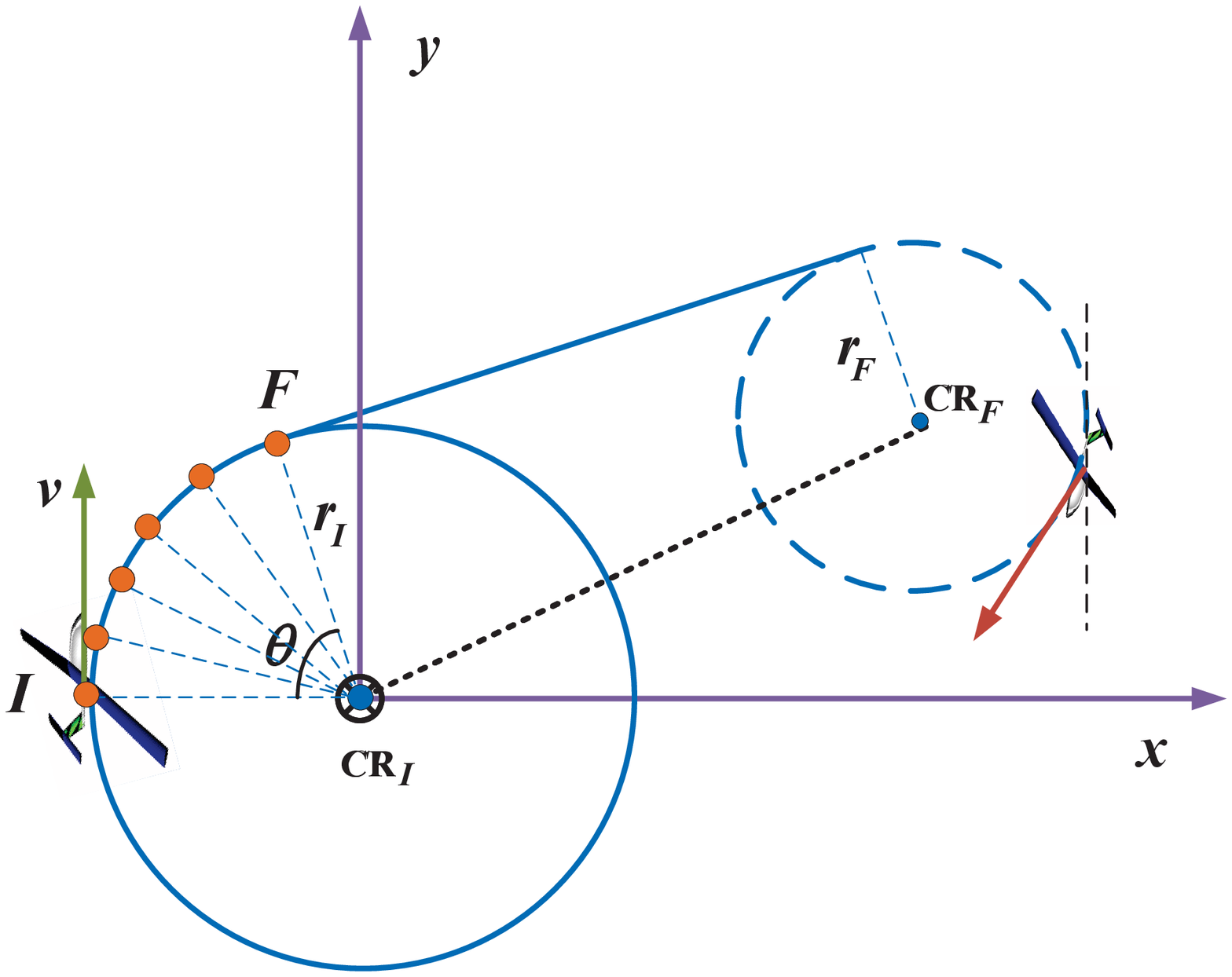}}
\caption{ The 3D and 2D illustrations of the fixed-wing UAV-enabled wireless network.}
\label{Fig5}
\end{figure}
Without loss of generality, we establish a 3D Cartesian coordinate system with UC as the origin $(0,0,0)$. Since DSTM is continuous and repetitive, we only need to observe and analyze it within the initial trajectory adjustment period $T$. Further, we divide $T$ into $N$ time slots, $\mathcal{N} = \{1,2,...,N\}$ and the length of each time slot is $\delta=\frac {T}{N}$. The initial and final flight radii in period $T$ are denoted as $r_{I}$ and $r_{F}$, respectively. In this paper, we aim to provide fair coverage for each MGU while ensuring flying security. The safe flight radius of the fixed-wing UAV shouldn't be less than 200 m \cite{FixSwarms}. Hence, we take the UAV flight radii $r_I=\max\{\bar{r}_1/2,200\}$ and $r_F=\max\{\bar{r}_N/2,200\}$, respectively, where $\bar{r}_1$ and $\bar{r}_N$ are the distribution radii of MGUs in time slot $n=1$ and $n=N$, respectively. Then, the UAV trajectory projected on the horizontal plane in time slot $n$ can be expressed as $ \vect{ q}[n]=\left(-r_{I}\cos[\frac {v}{r_{I}}n],r_{I}\sin[\frac {v}{r_{I}}n]\right), n\in\mathcal{N}$, where $v$ is the velocity of the UAV. In practical DSTM, the UAV needs to satisfy initial/final position constraints, which can be expressed as
\begin{equation}\label{e1}
\vect{q}[1]=\vect{q}_{I}, ~\vect{q}[N]=\vect{q}_{F},
\end{equation}
where $\vect{q}_{I}$ and $\vect{q}_{F}$ represent the UAV's destined initial and final locations, respectively. In addition, the maximal and minimal flying speeds of the UAV are denoted as $V_{\textmd{max}}$ and $V_{\textmd{min}}$, respectively. Accordingly, the maximum and minimum flight distance of each time slot can be expressed as $S_{\textmd{max}}=V_{\textmd{max}}\delta$ and $S_{\textmd{min}}=V_{\textmd{min}}\delta$, respectively. We denote $\wideparen{\vect{q}[n]\vect{q}[n+1]}$ as the arc length of the UAV moving from time slot $n$ to time slot $n+1$. The specific calculation of $\wideparen{\vect{q}[n]\vect{q}[n+1]}$ can be referred to Appendix A. Thus, the UAV trajectory is restricted by
\begin{equation}\label{2}
S_{\textmd{min}}\leq\wideparen{\vect{q}[n]\vect{q}[n+1]}\leq S_{\textmd{max}},  ~\forall n.
\end{equation}

On the 2D projection plane shown in Fig. 1 (b), the initial circle center coordinate, denoted as $CR_{I}$, is $(0,0)$. The final circle center coordinate $CR_{F}=(x_{RF},y_{RF})$ can be obtained according to user mobility prediction within period $T$. The angle $\theta\in[0,\pi)$ is the included angle formed by the line between the initial point $I$ and tangent point $F$ and center $CR_{I}$. Please refer to Appendix A for specific calculation of $\theta$. Due to the curvature constraint of the fixed-wing UAV during flight, the trajectory can only be switched at the tangent point $F$. Therefore, the flight velocity $v$ of the fixed-wing UAV needs to meet the following limitations:
\begin{equation}\label{3}
Cir=\frac {Tv-\theta r_{I}}{2\pi r_{I}}\in \mathbb{Z}_{0},
\end{equation}
\begin{equation}\label{4}
V_{\textmd{min}}\leq v \leq V_{\textmd{max}},
\end{equation}
where $Cir$ denotes the number of detour circles from initial point $I$ back to initial point $I$, which is a non-negative integer. $\mathbb{Z}_{0}$ is the non-negative set.

%\begin{figure}[h]
%\centering
%\includegraphics[width=1\columnwidth]{Fig2.eps}
%\caption{Fixed-wing UAV trajectory switching (planform).}
%\label{Fig2}
%\end{figure}

We consider that the fixed-wing UAV-enabled wireless network consists of $K$ MGUs denoted by the set $\mathcal{K} = \{1,2,...,K\}$. The initial locations of $K$ MGUs are randomly distributed, and the horizontal position are denoted as $ \vect{ s}_{k}[n]=\left(x_{k}[n],y_{k}[n]\right), k\in\mathcal{K}$. The UAV is assumed to fly at a fixed altitude $H$. Then, the distance from the UAV to user $k$ in time slot $n$ can be written as
\begin{equation}\label{5}
d_{k}[n]=\sqrt{H^{2}+\left\|\vect{q}[n]-\vect{s}_{k}[n]\right\|^{2}}, ~\forall n,k,
\end{equation}
where $\|\cdot\|$ represents Euclidean norm.

Generally, the channels between the UAV and users are dominated by the LOS links \cite{Tsinghua}. In other words, the communication quality depends on the distance of the UAV to the user. We assume that the Doppler effects induced by both mobility of the UAV and MGUs can be perfectly compensated at the receivers. Thus, the channel between the UAV and user $k$ during the time slot $n$ follows the free-space path-loss model, which can be modeled as
\begin{equation}\label{6}
h_{k}[n]=\frac{\rho_{0}}{d_{k}^{2}[n]}=\frac{\rho_{0}}{H^{2}+\left\|\vect{q}[n]-\vect{s}_{k}[n]\right\|^{2}}, ~\forall n,k,
\end{equation}
where $\rho_{0}$ denotes the channel power gain at the reference distance $d_{0}=1$ m.

In this paper, the OFDMA protocol is applied for the UAV BS to serve multiple users. The total bandwidth of the UAV-enabled system is $B_{\textmd{max}}$ and the peak transmission power of the UAV is denoted as $P_{\textmd{max}}$. $b_{k}[n]$ and $p_{k}[n]$ represent the bandwidth allocation and the downlink transmission power of the UAV to user $k$ in time slot $n$, respectively. Therefore, the constraints of bandwidth and transmission power can be expressed as
\begin{equation}\label{7}
\sum_{k = 1}^{K} \alpha_{k}[n]b_{k}[n]\leq B_{\textmd{max}}, ~\forall n,
\end{equation}
\begin{equation}\label{8}
\sum_{k = 1}^{K} \alpha_{k}[n]p_{k}[n]\leq P_{\textmd{max}}, ~\forall n,
\end{equation}
where $\alpha_{k}[n]$ is a binary variable used to distinguish different scheduling statuses. $\alpha_{k}[n]=1$ indicates user $k$ is scheduled by UAV BS in time slot $n$, otherwise, $\alpha_{k}[n]=0$. If user $k$ is served by UAV BS in time slot $n$,  i.e., $\alpha_{k}[n]=1$, the corresponding received signal-to-noise ratio (SNR) at user $k$ from UAV is given by
\begin{equation}\label{9}
\gamma_{k}[n]=\frac{p_{k}[n]h_{k}[n]}{N_{0}b_{k}[n]}, ~\forall n,k,
\end{equation}
where $N_{0}$ represents the power spectral density of the additive white Gaussian noise (AWGN) at the receivers. Accordingly, the achievable rate of user $k$ in time slot $n$, denoted by $R_{k}[n]$ in bits/second (bps), can be expressed as
\begin{equation}\label{10}
%R_{k}[n]=b_{k}[n]\log_{2}(1+\gamma_{k}[n])=b_{k}[n]\log_{2}(1+\frac{\rho_{0}p_{k}[n]}{{N_{0}b_{k}[n]H^{2}+\|\vect{q}[n]-\vect{s}_{k}[n]\|^{2}}),\forall n,k
R_{k}[n]=b_{k}[n]\log_{2}(1+\gamma_{k}[n]), ~\forall n,k.
\end{equation}

However, due to the dual mobility of the UAV and ground MGUs, the air-to-ground channel condition is not stable. To guarantee the QoS of MGUs, the data rate cannot be lower than a predetermined threshold $\gamma^{\textmd{th}}$. Therefore, we have
\begin{equation}\label{11}
R_{k}[n]\geq \alpha_{k}[n]\gamma^{\textmd{th}}, ~\forall n,k.
\end{equation}

As a result, the average achievable throughput of user $k$ over $N$ time slots is the function of $\alpha_{k}[n]$, $b_{k}[n]$, $p_{k}[n]$, and $\vect{q}[n]$, which is given by
\begin{equation}\label{12}
\bar{R}_{k}(\alpha_{k}[n],b_{k}[n],p_{k}[n],\vect{q}[n])=\frac{1}{N}\sum_{n=1}^{N}\alpha_{k}[n]R_{k}[n], ~\forall k.
\end{equation}

\subsection{Problem Formulation}
In this paper, we investigate the joint optimization problem of user scheduling, bandwidth allocation, power control as well as UAV trajectory control to maximize the minimum average throughput among users. For notational simplicity, let user scheduling $\boldsymbol {\alpha}=\{\alpha_{k}[n],\forall k,n\}$, bandwidth allocation $\vect B=\{b_{k}[n],\forall k,n\}$, power control $\vect P=\{p_{k}[n],\forall k,n\}$, and UAV trajectory control $\vect Q=\{\vect{q}[n],\forall n\}$. Define $\eta(\boldsymbol {\alpha},\vect B,\vect P,\vect Q)\triangleq\min\limits_{k\in\mathcal{K}}\bar{R}_{k}$ as a function of $\boldsymbol {\alpha}$, $\vect B$, $\vect P$, and $\vect Q$. The optimization problem can be formulated as
\begin{subequations} \label{13}
\begin{align}
%{\mathbf{P1}:~} & \mathrm{variable}~~ a_{uk}, ~b_{ut}, ~{P_{uk}(t)},~\bm \Phi(t), \forall u,k,t
{~} &  \mathrm{\max\limits_{\eta,\boldsymbol {\alpha},\vect B,\vect P,\vect Q}}~~~~~~~     \eta \label{OBJ}% \sum_{u\in\mathcal{U}_{s}} R_{u},  \label{optP}
\\& \mathrm{s.t.}~~~~~~\vect{q}[1]=\vect{q}_{I}, ~ \vect{q}[N]=\vect{q}_{F},  \label{13b}
\\& ~~~~~~\quad S_{\textmd{min}}\leq\wideparen{\vect{q}[n]\vect{q}[n+1]}\leq S_{\textmd{max}} , ~\forall n, \label{13c}
%\\& ~~~~~~~~~~~ \eta_{i}\geq \eta_{0}, \forall i\in \mathcal{U}_{I} \label{optiot}
\\&      ~~~~~~\quad \alpha_{k}[n]\in\{0,1\},~\forall k,n, \label{13d}
\\&       ~~~~~~\quad \frac{1}{N}\sum_{n=1}^{N}\alpha_{k}[n]R_{k}[n]\geq\eta,~\forall k,\label{13e}
%\\&      ~~~~~~\quad\quad    \sum_{k=1}^{K}\sum_{t=1}^{T} \rho_{kt}^{{u}}\geq1, \forall u\in\mathcal{U},  \label{optf}
\\&      ~~~~~~\quad    R_{k}[n]\geq\alpha_{k}[n]\gamma^{\textmd{th}},~\forall k,n,  \label{13f}
\\&      ~~~~~~\quad    \sum_{k = 1}^{K} \alpha_{k}[n]b_{k}[n]\leq B_{\textmd{max}}, ~\forall n,  \label{13g}
\\&      ~~~~~~\quad    0\leq b_{k}[n]\leq B_{\textmd{max}}, ~\forall k,n,  \label{13h}
\\&      ~~~~~~\quad    \sum_{k = 1}^{K} \alpha_{k}[n]p_{k}[n]\leq P_{\textmd{max}}, ~\forall n,  \label{13i}
\\&      ~~~~~~\quad   0\leq p_{k}[n]\leq P_{\textmd{max}}, ~\forall k,n.  \label{13j}
%\\&      ~~~~~~~~~~    |U|\geq N_{\mathrm{min}}^{\mathrm{users}},
\end{align}
\end{subequations}

Constraint (\ref{13b}) represents the initial and the final positions of the UAV. The flying position constraint of the UAV during period $T$ is shown in (\ref{13c}). Constraint (\ref{13d}) is the binary constraint to indicate whether the user $k$ is scheduled in time slot $n$. Constraint (\ref{13e}) is imposed to guarantee an average minimum rate $\eta$ for each user within period $T$. To ensure the QoS, the minimum communication rate constraint is given in (\ref{13f}). Bandwidth allocation and power control constraints are shown in (\ref{13g})-(\ref{13h}) and (\ref{13i})-(\ref{13j}), respectively.

It can be seen that problem (\ref{13}) is a mixed-integer non-convex optimization problem, which is difficult to obtain the optimal solution in general. There are three challenges to solve problem (\ref{13}). Firstly, multivariate variables are coupled and the objective function $\eta$ is non-concave. Secondly, the user scheduling $\alpha_{k}[n]$ in constraints (\ref{13d})-(\ref{13g}), and (\ref{13i}) is a binary variable, which cannot be solved directly. Finally, even though with the fixed user scheduling $\boldsymbol {\alpha}$, objective function $\eta$ and constraints (\ref{13e})-(\ref{13f}) are not jointly concave with respect to the optimization variables $\vect B$, $\vect P$, and $\vect Q$. To solve problem (\ref{13}), we decompose the original problem into three optimization subproblems and propose an efficient iterative algorithm.

\section{Joint User Scheduling, Resource Allocation, and Trajectory Control Algorithm Design}\label{section:Joint Optimization of User Scheduling, Resource Allocation, and Trajectory Control Algorithm Design}
In this section, we maximize the minimum average throughput among users by iteratively optimizing user scheduling, joint bandwidth and power allocation, and UAV trajectory control. More specifically, given the bandwidth allocation $\vect B$, transmission power $\vect P$, and UAV trajectory control $\vect Q$, the user scheduling $\boldsymbol {\alpha}$ is first optimized by solving a linear programming (LP) problem. Then, given the user scheduling $\boldsymbol {\alpha}$ and UAV trajectory $\vect Q$, the objective function $\eta$ is jointly concave with respect to $\vect P$ and $\vect B$, which can be tackled by applying Lagrange duality. Finally, with fixed user scheduling and resource allocation (i.e., $\boldsymbol {\alpha},\vect B,\vect P$), we optimize the UAV trajectory $\vect Q$ by applying SCA.

\subsection{User Scheduling Optimization}
Since we assume that the UAV BS employs OFDMA to serve MGUs, user scheduling optimization will be performed in each time slot. In order to make the optimization problem (\ref{13}) tractable, we first relax the binary variable $\boldsymbol {\alpha}$ into a continuous variable as follows
\begin{equation}\label{14}
0\leq \hat{\alpha}_{k}[n]\leq 1, ~\forall k,n.
\end{equation}

We denote $\hat{\boldsymbol{\alpha}}=\{\hat{\alpha}_{k}[n],\forall k,n\}$ as a set of user scheduling variables. With fixed bandwidth allocation, power control, and UAV trajectory control (i.e., $\vect B,\vect P,\vect Q$), problem (\ref{13}) can be rewritten as
\begin{subequations} \label{15}
\begin{align}
{~} &  \mathrm{\max\limits_{\eta,\hat{\boldsymbol {\alpha}}}}~~~~~~~~~~~~     \eta \label{optOBJ}% \sum_{u\in\mathcal{U}_{s}} R_{u},  \label{optP}
\\& \mathrm{s.t.}~~~~ (\textmd{\ref{13e}})-(\textmd{\ref{13g}}), (\textmd{\ref{13i}}),(\ref{14}).\label{15b}
\end{align}
\end{subequations}

It is straightforward to observe that problem (\ref{15}) is a standard LP problem, which can be solved efficiently by using the optimization toolbox CVX \cite{CVX}.

\subsection{Joint Bandwidth Allocation and Power Control}

In this subsection, we jointly optimize the bandwidth allocation $\vect B$ and power control $\vect P$ by assuming that user scheduling $\hat{\boldsymbol{\alpha}}$ and UAV trajectory $\vect Q$ are fixed. Thus, the resource allocation optimization in problem (\ref{13}) can be reformulated as
\begin{subequations} \label{16}
\begin{align}
%{\mathbf{P1}:~} & \mathrm{variable}~~ a_{uk}, ~b_{ut}, ~{P_{uk}(t)},~\bm \Phi(t), \forall u,k,t
{~} &  \mathrm{\max\limits_{\eta,\vect B,\vect P}}~~~~~~~~~~\eta \label{optOBJ}% \sum_{u\in\mathcal{U}_{s}} R_{u},  \label{optP}
\\& \mathrm{s.t.}~\frac{1}{N}\sum_{n=1}^{N}\hat{\alpha}_{k}[n]b_{k}[n]\log_{2}\left(1+\frac{p_{k}[n]g_{k}[n]}{b_{k}[n]}\right)\geq\eta,~\forall k,  \label{16b}
\\& ~~~~~~ b_{k}[n]\log_{2}\left(1+\frac{p_{k}[n]g_{k}[n]}{b_{k}[n]}\right)\geq\hat{\alpha}_{k}[n]\gamma^{\textmd{th}}, ~\forall k,n, \label{16c}
\\&      ~~~~~~\quad    \sum_{k = 1}^{K} \hat{\alpha}_{k}[n]b_{k}[n]\leq B_{\textmd{max}}, ~\forall n,  \label{16d}
\\&      ~~~~~~\quad    0\leq b_{k}[n]\leq B_{\textmd{max}}, ~\forall k,n,  \label{16e}
\\&      ~~~~~~\quad    \sum_{k = 1}^{K} \hat{\alpha}_{k}[n]p_{k}[n]\leq P_{\textmd{max}}, ~\forall n,  \label{16f}
\\&      ~~~~~~\quad   0\leq p_{k}[n]\leq P_{\textmd{max}}, ~\forall k,n,  \label{16g}
\end{align}
\end{subequations}
where $g_{k}[n]\triangleq\frac{\rho_{0}}{N_{0}d_{k}^{2}[n]}$, $\forall k,n$. Obviously, (\ref{16d})-(\ref{16g}) are all affine constraints. We only need to verify the convexity of the objective function $\eta$ and constraints (\ref{16b})-(\ref{16c}) with respect to $b_{k}[n]$ and $p_{k}[n]$ to derive the optimal solution of problem (\ref{16}).

\emph{Lemma 1:} For any constant $a\geq0$, the function $\psi(x,y)\triangleq x\log_{2}\left(1+\frac{ay}{x}\right)$ is jointly concave with respect to $x\geq0$ and $y\geq0$.

\emph{Proof:} Please refer to Appendix B.

According to Lemma 1, problem (\ref{16}) is a convex optimization problem since the function $b_{k}[n]\log_{2}\left(1+\frac{p_{k}[n]g_{k}[n]}{b_{k}[n]}\right)$ is jointly concave with respect to $b_{k}[n]$ and $p_{k}[n]$. Besides, problem (\ref{16}) satisfies the Slater's constraint, so the strong duality holds \cite{LiRuide}. In other words, the duality gap between problem (\ref{16}) and its dual problem is zero and the optimization can be achieved by solving the Lagrange duality. The Lagrangian function of problem (\ref{16}) can be written as
\begin{equation}\label{17}
    \begin{aligned}
    &\mathcal{L}(\eta,\vect B,\vect P,\boldsymbol{\mu},\boldsymbol{\beta},\boldsymbol{\xi},\boldsymbol{\varpi})=\eta-\sum_{k=1}^{K}\mu_{k}\left(N\eta-\sum_{n=1}^{N}\hat{\alpha}_{k}[n]b_{k}[n]\log_{2}\left(1+\frac{p_{k}[n]g_{k}[n]}{b_{k}[n]}\right)\right)\\
    &-\sum_{k=1}^{K}\sum_{n=1}^{N}\beta_{k,n}\left(\hat{\alpha}_{k}[n]\gamma^{\textmd{th}}-b_{k}[n]\log_{2}\left(1+\frac{p_{k}[n]g_{k}[n]}{b_{k}[n]}\right)\right)-\sum_{n=1}^{N}\xi_{n}\left(\sum_{k=1}^{K}\hat{\alpha}_{k}[n]b_{k}[n]-B_{\textmd{max}}\right)\\
    &-\sum_{n=1}^{N}\varpi_{n}\left(\sum_{k=1}^{K}\hat{\alpha}_{k}[n]p_{k}[n]-P_{\textmd{max}}\right),
    \end{aligned}
    \end{equation}
where $\boldsymbol{\mu}=\{\mu_{k},\forall k\}$, $\boldsymbol{\beta}=\{\beta_{k,n},\forall k,n\}$, $\boldsymbol{\xi}=\{\xi_{n},\forall n\}$, and $\boldsymbol{\varpi}=\{\varpi_{n},\forall n\}$ denote the non-negative Lagrange multiplier vectors for constraints (\ref{16b}), (\ref{16c}), (\ref{16d}), and (\ref{16f}), respectively. Constraints (\ref{16e}) and (\ref{16g}) will be absorbed into the Karush-Kuhn-Tucker (KKT) conditions \cite{LayerNg} when deriving the optimal solution of resource allocation. Hence, the dual problem of problem (\ref{16}) can be expressed as
\begin{equation} \label{18}
\begin{aligned}
{~} & \mathrm{\min\limits_{\boldsymbol{\mu},\boldsymbol{\beta},\boldsymbol{\xi},\boldsymbol{\varpi} \geq 0}\max\limits_{\eta,\vect B,\vect P}~\mathcal{L}(\eta,\vect B,\vect P,\boldsymbol{\mu},\boldsymbol{\beta},\boldsymbol{\xi},\boldsymbol{\varpi})}.
\end{aligned}
\end{equation}

Next, we solve the dual problem iteratively by decomposing it into two layers \cite{Wuqing,LiRuide}.

\subsubsection{Solution of Layer 1 (Power Control and Bandwidth Allocation)}

By dual decomposition, we first solve the Lagrange dual function with fixed dual variables $\boldsymbol{\mu}$, $\boldsymbol{\beta}$, $\boldsymbol{\xi}$, and $\boldsymbol{\varpi}$, i.e.,
\begin{equation} \label{19}
\begin{aligned}
{~} & \mathrm{\max\limits_{\eta,\vect B,\vect P}~\mathcal{L}(\eta,\vect B,\vect P,\boldsymbol{\mu},\boldsymbol{\beta},\boldsymbol{\xi},\boldsymbol{\varpi})}.
\end{aligned}
\end{equation}

Since the objective function $\eta$ represents the minimum average throughput of users, for simplicity, we set $\eta^*=0$ as the initial optimal solution to obtain the dual function (19). With the given dual variables $\boldsymbol{\mu}$, $\boldsymbol{\beta}$, $\boldsymbol{\xi}$, and $\boldsymbol{\varpi}$, problem (\ref{19}) is jointly concave with respect to $p_{k}[n]$ and $b_{k}[n]$. According to the KKT conditions, the optimal power allocation for user $k$ in time slot $n$, denoted as $p_{k}^{*}[n]$, is given by
\begin{equation} \label{20}
p_{k}^{*}[n]=b_{k}[n]\left[\frac{\mu_{k}+N\beta_{k,n}}{\varpi_{n}N\ln2}-\frac{1}{g_{k}[n]}\right]^{+},
\end{equation}
where $[p]^{+}$ represents $\max\{p,0\}$. The power control in (\ref{20}) follows the multi-level water-filling policy \cite{Wuqing}. Let  $\tilde{p}_{k}[n]\triangleq\frac{p_{k}^{*}[n]}{b_{k}[n]}=\left[\frac{\mu_{k}+N\beta_{k,n}}{\varpi_{n}N\ln2}-\frac{1}{g_{k}[n]}\right]^{+}$, which is uniquely determined with fixed $\boldsymbol{\mu}$, $\boldsymbol{\beta}$, and $\boldsymbol{\varpi}$. By substituting $\tilde{p}_{k}[n]$ back into problem (\ref{19}), we have
\begin{subequations}\label{21}
\begin{align}
&\mathrm{\max\limits_{\vect B}}~~~\sum_{k=1}^{K}\sum_{n=1}^{N}f(\tilde{p}_{k})b_{k}[n]-\Gamma
\\& \mathrm{s.t.}~~~~~~~~~(\textmd{\ref{16e}}), \label{21b}
\end{align}
\end{subequations}
where
\begin{eqnarray} \label{22}
f(\tilde{p}_{k})=\frac{\mu_{k}+N\beta_{k,n}}{N}\hat{\alpha}_{k}[n]\log_{2}\left(1+g_{k}[n]\tilde{p}_{k}[n]\right)-\varpi_{n}\hat{\alpha}_{k}[n]\tilde{p}_{k}[n]-\xi_{n}\hat{\alpha}_{k}[n],
\end{eqnarray}
\begin{eqnarray} \label{23}
\Gamma=\sum_{k=1}^{K}\sum_{n=1}^{N}\beta_{k,n}\hat{\alpha}_{k}[n]\gamma^{\textmd{th}}-\sum_{n=1}^{N}\xi_{n}B_{\textmd{max}}-\sum_{n=1}^{N}\varpi_nP_{\textmd{max}}.
\end{eqnarray}

Obviously, problem (\ref{21}) is an LP problem about optimizing the variable $b_{k}[n]$. Therefore, the optimal bandwidth allocation for user $k$ in time slot $n$, denoted as $b_{k}^{*}[n]$, is given by
\begin{equation}\label{23}
b_{k}^{*}[n]=\left\{
\begin{array}{rcl}
B_{\textmd{max}}, &&  f(\tilde{p}_{k})>0,~~~~~~~~\forall k,n,\\
0, && \textmd{otherwise},
\end{array}\right.
\end{equation}
Note that we set $b_{k}^{*}[n]=0$ when $f(\tilde{p}_{k})=0$ holds, since the objective function value in problem (\ref{21}) is not affected by the value of $b_{k}^{*}[n]$.

\subsubsection{Solution of Layer 2 (Solving the Dual Problem (\ref{18}))}

After obtaining $\eta^{*}$, $\vect {B}^{*}$, and $\vect {P}^{*}$, we can update the Lagrange multipliers by applying the gradient method \cite{LiRuide}, because the dual function (18) is differentiable. Then, the gradient update equations of dual variables are given as follows
\begin{equation} \label{24}
\mu_{k}(m+1)=\left[\mu_{k}(m)-\zeta_1(m)\times\left(\frac{1}{N}\sum_{n=1}^{N}\hat{\alpha}_{k}[n]b_{k}[n]\log_{2}\left(1+\frac{p_{k}[n]g_{k}[n]}{b_{k}[n]}\right)-\eta\right)\right]^{+}, ~\forall k,
\end{equation}
\begin{equation} \label{25}
\beta_{k,n}(m+1)=\left[\beta_{k,n}(m)-\zeta_2(m)\times\left(b_{k}[n]\log_{2}\left(1+\frac{p_{k}[n]g_{k}[n]}{b_{k}[n]}\right)-\hat{\alpha}_{k}[n]\gamma^{\textmd{th}}\right)\right]^{+}, ~\forall k,n,
\end{equation}
\begin{equation} \label{26}
\xi_{n}(m+1)=\left[\xi_{n}(m)-\zeta_3(m)\times\left(B_{\textmd{max}}-\sum_{k=1}^{K}\hat{\alpha}_{k}[n]b_{k}[n]\right)\right]^+, ~\forall n,
\end{equation}
\begin{equation} \label{27}
\varpi_{n}(m+1)=\left[\varpi_n(m)-\zeta_4(m)\times\left(P_{\textmd{max}}-\sum_{k=1}^{K}\hat{\alpha}_{k}[n]p_{k}[n]\right)\right]^+, ~\forall n,
\end{equation}
where $m\geq0$ represents the iteration index and $\zeta_u(m)$, $u\in{1,2,3,4}$, are positive step sizes. In addition, the details such as selecting the step size and proving the convergence of the gradient method can be found in \cite{LayerNg,Subg}.

The optimal solution that maximizes the Lagrangian function is equal to the optimal primal solution if and only if the solution is feasible and unique \cite{Convex}. However, from the solution analysis of Layer 1, the optimal values of $\eta^{*}$ and $\vect B^{*}$ are not unique because of the initial set $\eta^*=0$ and $f(\tilde{p}_{k})=0$. Therefore, additional steps are needed to further confirm the values of $\eta^{*}$ and $\vect B^{*}$. We denote $\tilde{p}_{k}^{*}[n]=\frac{p_{k}^{*}[n]}{b_{k}^{*}[n]}$ as the optimal power spectrum density. Since $\tilde{p}_{k}^{*}[n]$ can be uniquely obtained from (\ref{20}), we substitute it into the original problem (\ref{16}) to obtain an LP problem with respect to $\vect B$ and $\eta$, which can be solved by CVX \cite{CVX}. Finally, the corresponding power control $\vect {P}^{*}$ can be calculated by $p_{k}^{*}[n]=\tilde{p}_{k}^{*}[n]b_{k}^{*}[n]$ with the optimal bandwidth allocation $\vect {B}^{*}$.

\subsection{UAV Trajectory Optimization}

Given any feasible user scheduling and resource allocation (i.e., $\boldsymbol{\hat{\alpha}}, \vect B, \vect P$), problem (13) can be represented as
\begin{subequations} \label{}
\begin{align}
%{\mathbf{P1}:~} & \mathrm{variable}~~ a_{uk}, ~b_{ut}, ~{P_{uk}(t)},~\bm \Phi(t), \forall u,k,t
{~} &  \mathrm{\max\limits_{\eta,\vect Q}}~~~~~~~~~~     \eta \label{OBJ}% \sum_{u\in\mathcal{U}_{s}} R_{u},  \label{optP}
\\& \mathrm{s.t.}~~~~~~\vect{q}[1]=\vect{q}_{I}, ~ \vect{q}[N]=\vect{q}_{F},  \label{b}
\\& ~~~~~~\quad S_{\textmd{min}}\leq\wideparen{\vect{q}[n]\vect{q}[n+1]}\leq S_{\textmd{max}} , ~\forall n, \label{c}
\\&       ~~~~~~\quad \frac{1}{N}\sum_{n=1}^{N}\hat{\alpha}_{k}[n]R_{k}[n]\geq\eta,~\forall k,\label{e}
%\\&      ~~~~~~\quad\quad    \sum_{k=1}^{K}\sum_{t=1}^{T} \rho_{kt}^{{u}}\geq1, \forall u\in\mathcal{U},  \label{optf}
\\&      ~~~~~~\quad    R_{k}[n]\geq\hat{\alpha}_{k}[n]\gamma^{\textmd{th}},~\forall k,n.  \label{f}
\end{align}
\end{subequations}

The optimization problem (29) is non-convex due to constraints (29c)-(29e), where neither $\wideparen{\vect{q}[n]\vect{q}[n+1]}$ nor $R_{k}[n]$ is concave with respect to $\vect {q}[n]$. In general, there is no efficient method to tackle such a non-convex problem. However, we find that the location of the UAV in time slot $n$ is $ \vect{q}[n]=\left(-r_{I}\cos[\frac {v}{r_{I}}n],r_{I}\sin[\frac {v}{r_{I}}n]\right), \forall n$, which is a function of the UAV's velocity $v$. Thus, we can obtain the optimal trajectory by optimizing the flying velocity of the UAV. With fixed $\boldsymbol{\hat{\alpha}}$, $\vect B$, and $\vect P$, the average achievable throughput $\bar{R}_{k}(v)$ of user $k$ over $N$ time slots is given by
\begin{equation}\label{28}
    \begin{aligned}
    &\bar{R}_{k}(v)=\frac{1}{N}\sum_{n=1}^{N}\hat{\alpha}_{k}[n]b_{k}[n]\log_{2}\Bigg(1+\frac{\rho_{0}p_{k}[n]/\left(N_{0}b_{k}[n]\right)}{H^{2}+\left(-r_{I}\cos[\frac {v}{r_{I}}n]-x_{k}[n]\right)^{2}+\left(r_{I}\sin[\frac {v}{r_{I}}n]-y_{k}[n]\right)^{2}}\Bigg)\\
    &=\frac{1}{N}\sum_{n=1}^{N}\vartheta_{k}[n]\log_{2}\left(1+\frac{\chi_{k}[n]}{\lambda_{k}[n]+\varsigma_{k}[n]\sin\left[\frac{v}{r_{I}}n-\phi_{k}[n]\right]}\right)\\
    &=\frac{1}{N}\sum_{n=1}^{N}\vartheta_{k}[n]\Bigg(\log_{2}\left(\chi_{k}[n]+\lambda_{k}[n]+\varsigma_{k}[n]\sin\left[\frac{v}{r_{I}}n-\phi_{k}[n]\right]\right)\\
    &-\log_{2}\left(\lambda_{k}[n]+\varsigma_{k}[n]\sin\left[\frac{v}{r_{I}}n-\phi_{k}[n]\right]\right)\Bigg),
    \end{aligned}
\end{equation}
where $\vartheta_k[n]=\hat{\alpha}_{k}[n]b_{k}[n],\forall k,n$, $\chi_{k}[n]=\rho_{0}p_{k}[n]/\left(N_{0}b_{k}[n]\right),\forall k,n$, $\lambda_{k}[n]=H^{2}+r_{I}^{2}+x_{k}^{2}[n]+y_{k}^{2}[n],\forall k,n$, $\varsigma_{k}[n]=2r_{I}\sqrt{x_{k}^{2}[n]+y_{k}^{2}[n]},\forall k,n$, and $\phi_{k}[n]=\tan^{-1}\left(y_{k}[n]/x_{k}[n]\right),\forall k,n$, are all constants. Thus, problem (29) can be reformulated to optimize $\eta$ and the UAV's velocity $v$, i.e.,
%\begin{figure*}[hb] %hb代表放在文章底部，%ht 为放在文章顶部
% 	\centering
% 	\begin{equation}\label{32}
%    \begin{aligned}
%    &\bar{R}_{k}(v)=\frac{1}{N}\sum_{n=1}^{N}\hat{\alpha}_{k}[n]b_{k}[n]\log_{2}\Bigg(1+\frac{\rho_{0}p_{k}[n]/\left(N_{0}b_{k}[n]\right)}{H^{2}+\left(-r_{I}\cos[\frac {v}{r_{I}}n]-x_{k}[n]\right)^{2}+\left(r_{I}\sin[\frac {v}{r_{I}}n]-y_{k}[n]\right)^{2}}\Bigg)\\
%    &=\frac{1}{N}\sum_{n=1}^{N}\hat{\alpha}_{k}[n]b_{k}[n]\log_{2}\left(1+\frac{\chi_{k}[n]}{\lambda_{k}[n]+\varsigma_{k}[n]\sin\left[\frac{v}{r_{I}}n-\phi_{k}[n]\right]}\right)\\
%    &=\frac{1}{N}\sum_{n=1}^{N}\hat{\alpha}_{k}[n]b_{k}[n]\Bigg(\log_{2}\left(\chi_{k}[n]+\lambda_{k}[n]+\varsigma_{k}[n]\sin\left[\frac{v}{r_{I}}n-\phi_{k}[n]\right]\right)\\
%    &-\log_{2}\left(\lambda_{k}[n]+\varsigma_{k}[n]\sin\left[\frac{v}{r_{I}}n-\phi_{k}[n]\right]\right)\Bigg)
%    \end{aligned}
%    \end{equation}
%
% \end{figure*}
\begin{subequations} \label{29}
\begin{align}
%{\mathbf{P1}:~} & \mathrm{variable}~~ a_{uk}, ~b_{ut}, ~{P_{uk}(t)},~\bm \Phi(t), \forall u,k,t
{~} &  \mathrm{\max\limits_{\eta,v}}~~~~~~~~     \eta \label{optOBJ}% \sum_{u\in\mathcal{U}_{s}} R_{u},  \label{optP}
\\& \mathrm{s.t.}~~~~~\frac {Tv-\theta r_{I}}{2\pi r_{I}}\in \mathbb{Z}_{0}, \label{29b}
\\& ~~~~~~\quad V_{\textmd{min}}\leq v\leq V_{\textmd{max}},  \label{29c}
\\& ~~~~~~\quad       %\frac{1}{N}\sum_{n=1}^{N}\hat{\alpha}_{k}[n]b_{k}[n]\log_{2}(1+\frac{\chi_{k}[n]}{\psi_{k}[n]+\varsigma_{k}[n]\sin[\frac{v}{r_{1}}n-\phi_{k}[n]]})\geq\eta,\forall k, \label{optd}
\bar{R}_k(v)\geq\eta, ~\forall k, \label{29d}
\\&    b_{k}[n]\log_{2}\left(1+\frac{\chi_{k}[n]}{\lambda_{k}[n]+\varsigma_{k}[n]\sin\left[\frac{v}{r_{I}}n-\phi_{k}[n]\right]}\right)\geq\hat{\alpha}_{k}[n]\gamma^{\textmd{th}}, ~\forall k,n. \label{29e}
\end{align}
\end{subequations}

Constraint (31b) represents that the final position of the UAV must be  at the tangent point $F$. The flying velocity constraint of the UAV during period $T$ is shown in constraint (31c). Constraint (31d) is to guarantee an average minimum rate $\eta$ for each user within period $T$. Constraint (31e) is imposed to ensure the minimum communication rate of user $k$ in each time slot $n$. Although the velocity $v$ is decoupled from other optimization variables (i.e., $\boldsymbol{\hat{\alpha}}, \vect B, \vect P$), problem (\ref{29}) is still non-convex due to the following reasons. On the one hand, the velocity $v$ in constraint (\ref{29b}) is discrete. On the other hand, the objective function $\eta$ and constraints (\ref{29d})-(\ref{29e}) are neither convex nor concave with respect to $v$. To tackle the non-convex problem (\ref{29}), we adopt the variable substitution and SCA \cite{SpaceAirGround} to further transform problem (\ref{29}) into a convex optimization problem. The auxiliary variable is introduced, which is given by
\begin{equation}\label{30}
{X}_{k}[n]=\sin\left[\frac{v}{r_{I}}n-\phi_{k}[n]\right], ~\forall k,n,
\end{equation}
where ${X}_{k}[n]\in [-1,1],\forall k,n$ disregarding the constraints of velocity $v$ in (\ref{29b})-(\ref{29c}). Define $\mathcal{F}_{k}^{1}[n]\triangleq\log_{2}\left(\chi_{k}[n]+\lambda_{k}[n]+\varsigma_{k}[n]{X}_{k}[n]\right), \forall k,n$ and $\mathcal{F}_{k}^{2}[n]\triangleq\log_{2}\left(\lambda_{k}[n]+\varsigma_{k}[n]{X}_{k}[n]\right), \forall k,n$. We denote $\mathcal{F}({X}_{k}[n])=\mathcal{F}_{k}^{1}[n]-\mathcal{F}_{k}^{2}[n],\forall k,n$. Obviously, both $\mathcal{F}_{k}^{1}[n]$ and $\mathcal{F}_{k}^{2}[n]$ are all concave functions about ${X}_{k}[n]$. Therefore, $\mathcal{F}({X}_{k}[n])$ is a difference of convex (DC) programming problem \cite{SpaceAirGround}, and its concavity and convexity is uncertain. In order to tackle the DC problem, we apply SCA to approximate $\mathcal{F}_{k}^{2}[n]$ in each iteration. That is to say, the first-order Taylor expansion is used to approximate $\mathcal{F}_{k}^{2}[n]$ to a linear function with a given local point ${X}_{k}^{i}[n]$ in the $i$-th iteration, i.e.,
\begin{equation}\label{31}
\begin{aligned}
&\log_{2}\left(\lambda_{k}[n]+\varsigma_{k}[n]{X}_{k}[n]\right)\leq\log_{2}\left(\lambda_{k}[n]+\varsigma_{k}[n]{X}_{k}^{i}[n]\right)+\frac{\varsigma_{k}[n]}{\left(\lambda_{k}[n]+\varsigma_{k}[n]{X}_{k}^{i}[n]\right)\ln2}\\
&\left({X}_{k}[n]-{X}_{k}^{i}[n]\right).
\end{aligned}
\end{equation}

Combining with (\ref{31}), $\mathcal{F}({X}_{k}[n])$ can be further expressed as
\begin{equation}\label{32}
\begin{aligned}
&\mathcal{F}({X}_{k}[n])\geq \mathcal{F}_{k}^{1}[n]-\Bigg(\frac{\varsigma_{k}[n]}{\left(\lambda_{k}[n]+\varsigma_{k}[n]{X}_{k}^{i}[n]\right)\ln2}\big({X}_{k}[n]-{X}_{k}^{i}[n]\big)+\log_{2}\left(\lambda_{k}[n]+\varsigma_{k}[n]{X}_{k}^{i}[n]\right)\Bigg)\\
&\triangleq \mathcal{F}^{\textmd{lb},i}\left({X}_{k}[n]\right),
\end{aligned}
\end{equation}
where $\mathcal{F}^{\textmd{lb},i}({X}_{k}[n])$ is the lower bounds of $\mathcal{F}({X}_{k}[n])$ in the $i$-th iteration. We denote $\vect X=\{{X}_{k}[n], \forall k,n\}$ as the set of auxiliary variables. As a result, problem (\ref{29}) can be approximated as
\begin{subequations} \label{33}
\begin{align}
%{\mathbf{P1}:~} & \mathrm{variable}~~ a_{uk}, ~b_{ut}, ~{P_{uk}(t)},~\bm \Phi(t), \forall u,k,t
{~} &  \mathrm{\max\limits_{\vect X}}~~~~~~~     \frac{1}{N}\sum_{n=1}^{N}\vartheta_k[n]\mathcal{F}^{\textmd{lb},i}\left({X}_{k}[n]\right) \label{optOBJ}% \sum_{u\in\mathcal{U}_{s}} R_{u},  \label{optP}
\\& \mathrm{s.t.}~~~~~~-1\leq {X}_{k}[n]\leq 1, ~\forall k,n,  \label{33b}
%\\& ~~~~~~\quad \frac{1}{N}\sum_{n=1}^{N}\vartheta_{k}[n]\mathcal{F}^{\textmd{lb},i}\left({X}_{k}[n]\right)\geq \eta, ~\forall k,
\\& ~~~~~~\quad b_{k}[n]\mathcal{F}^{\textmd{lb},i}\left({X}_{k}[n]\right)\geq \hat{\alpha}_{k}[n]\gamma^{\textmd{th}}, ~\forall k,n. \label{33c}
\end{align}
\end{subequations}

With concave objective function and constraints, problem (\ref{33}) can be solved by applying the CVX \cite{CVX} to obtain the optimal solution $\vect{X}^{*}$. Since the velocity constraints (\ref{29b})-(\ref{29c}) are relaxed in problem (\ref{33}), there may be feasible solutions in the optimal solution $\vect{X}^{*}$ that do not conform to the original problem (\ref{29}). In addition, a particular value of $\vect{X}^{*}$ may correspond to multiple feasible solutions of $v$ in (\ref{30}). Therefore, we substitute $\vect{X}^{*}$ back into (\ref{29}). Combined with constraints (\ref{29b}) and (\ref{29c}), we can get the optimal value of the UAV's velocity, which is denoted as $v^{*}$. Further, the optimal trajectory $\vect Q^*$ can be obtained.

\subsection{Overall Algorithm Integration}

After obtaining the results of the above three subproblems, we propose an overall iterative algorithm to get the optimization results of the original problem (\ref{13}). More specifically, problem (\ref{13}) includes four optimization variables, namely user scheduling $\boldsymbol \alpha$, bandwidth allocation $\vect B$, power control $\vect P$, and UAV trajectory control $\vect Q$, which is converted to optimize UAV's velocity $v$ in problem (\ref{29}). In each iteration, we first optimize user scheduling $\boldsymbol \alpha$ by solving problem (\ref{15}) with fixed resource allocation $($i.e., $\vect B, \vect P)$ and UAV trajectory $\vect Q$. Then, given user scheduling $\boldsymbol \alpha$ and UAV trajectory $\vect Q$, the bandwidth allocation $\vect B$ and power control $\vect P$  can be obtained by solving problem (\ref{16}). Finally, with fixed user scheduling $\boldsymbol \alpha$ and resource allocation, the velocity $v$ of the UAV can be optimized by solving problem (\ref{29}), and the UAV trajectory $\vect Q$ can be further obtained. Moreover, the result from each iteration is used as the input for the next iteration until the convergence is achieved. The optimization algorithm for joint user scheduling, resource allocation, and UAV trajectory control is summarized in Algorithm 1.

\subsection{Convergence and Complexity Analysis}

Problem (\ref{13}) is decomposed into three subproblems. It is worth pointing out that for the user scheduling subproblem (\ref{15}) and the UAV trajectory control subproblem (29), we only optimally solve their approximate problems. To prove the convergence properties of Algorithm 1, we have the following analysis.

Define $\Omega\left(\boldsymbol{\alpha}^{l}, \vect{B}^{l}, \vect{P}^{l}, \vect Q^{l}\right)$ as the objective value of the original problem (\ref{13}) at the $l$-th iteration. In step 3 of Algorithm 1, since $\boldsymbol{\alpha}^{l+1}$ is suboptimal user scheduling of problem (\ref{15}) with the fixed $\vect{B}^{l}$, $\vect{P}^{l}$, $\vect Q^{l}$, we have
\begin{equation}\label{34}
\Omega\left(\boldsymbol{\alpha}^{l}, \vect{B}^{l}, \vect{P}^{l}, \vect Q^{l}\right)\leq \Omega\left(\boldsymbol{\alpha}^{l+1}, \vect{B}^{l}, \vect{P}^{l}, \vect Q^{l}\right).
\end{equation}

From step 4 to step 13 of Algorithm 1, $\vect{B}^{l+1}$ and $\vect{P}^{l+1}$ are the optimal bandwidth allocation and power control of problem (\ref{16}) with fixed $\boldsymbol{\alpha}^{l+1}$ and $\vect Q^{l}$, which follows that
\begin{equation}\label{35}
\Omega\left(\boldsymbol{\alpha}^{l+1}, \vect{B}^{l}, \vect{P}^{l}, \vect Q^{l}\right)\leq \Omega\left(\boldsymbol{\alpha}^{l+1}, \vect{B}^{l+1}, \vect{P}^{l+1}, \vect Q^{l}\right).
\end{equation}

\begin{figure}[!t]
\begin{flushleft}
\begin{spacing}{0.95}
\begin{tabular*}{\hsize}{@{}@{\extracolsep{\fill}}l@{}}
%  \arrayrulecolor{black}
  \hline
  \specialrule{0em}{1pt}{1pt}
  \textbf{Algorithm 1:}  Joint Optimization of User Scheduling, Resource Allocation, and Trajectory \\
  Control for Solving Problem (\ref{13})\\
  \specialrule{0em}{1pt}{1pt}
%  \arrayrulecolor{black}
  \hline
  \specialrule{0em}{1pt}{1pt}
  1:~Initialization: Initialize iterative index $l=0$, maximal iterations $L_{\textmd{max}}$, user scheduling $\boldsymbol{\alpha}^0$,\\
  ~~~bandwidth allocation $\vect{B}^{0}$, power allocation $\vect{P}^{0}$, UAV's trajectory $\vect{Q}^{0}$, and tolerance error $\epsilon$.\\
  2:  \bfseries repeat    \\
  3:~~~~Given $\vect{B}^{l}$, $\vect{P}^{l}$, and $\vect{Q}^{l}$, update user scheduling $\boldsymbol{\alpha}^{l+1}$ by solving problem (15);\\
  4:~~~~Initialize iterative index $m_\textmd{i}=0$, $m_\textmd{o}=0$, maximal iterations $M_{\textmd{inner}}$, $M_{\textmd{outer}}$, $\boldsymbol{\mu}$, $\boldsymbol{\beta}$, $\boldsymbol{\xi}$, and \\ ~~~~~~$\boldsymbol{\varpi}$;\\
  5:  ~~~\bfseries repeat    \\
  6:~~~~~~~Given $\boldsymbol{\alpha}^{l+1}$, $\vect{Q}^{l}$, $\boldsymbol{\mu}^{m_{\textmd{i}}}$, $\boldsymbol{\beta}^{m_{\textmd{i}}}$, $\boldsymbol{\xi}^{m_{\textmd{i}}}$, and $\boldsymbol{\varpi}^{m_{\textmd{i}}}$, obtain $\boldsymbol{\eta}^{*}$, $\vect{B}^{*}$, and $\vect{P}^{*}$ by solving problem \\
  ~~~~~~~~~(\ref{19});\\
  7:  ~~~~~~\bfseries repeat    \\
  8:~~~~~~~~~~Given $\boldsymbol{\eta}^{*}$, $\vect{B}^{*}$, and $\vect{P}^{*}$, update $\boldsymbol{\mu}^{m_{\textmd{i}}+1}$, $\boldsymbol{\beta}^{m_{\textmd{i}}+1}$, $\boldsymbol{\xi}^{m_{\textmd{i}}+1}$, and $\boldsymbol{\varpi}^{m_{\textmd{i}}+1}$ by applying the \\
  ~~~~~~~~~~~~gradient method to solve problem (18);\\
  9:~~~~~~~\textbf{until}~~convergence has been reached or $m_{\textmd{i}}\geq M_{\textmd{inner}}$;\\
  10:~~~~~~Update $m_{\textmd{o}}=m_{\textmd{o}}+1$;\\
  11:~~~\textbf{until}~~convergence has been reached or $m_{\textmd{o}}\geq M_{\textmd{outer}}$;\\
  12:~~~Given $\boldsymbol{\mu}^*$, $\boldsymbol{\beta}^*$, and $\boldsymbol{\varpi}^*$, compute $\tilde{p}_{k}^{*}[n]$ by using (\ref{20}), then determine the optimal $\boldsymbol{\eta}^{*}$, \\
  ~~~~~~~$\vect{B}^{*}$, and $\vect{P}^{*}$ by solving problem (16) with given $\tilde{p}_{k}^{*}[n]$;\\
  13:~~~Set $\vect B^{l+1}\leftarrow\vect{B}^{*}$, $\vect P^{l+1}\leftarrow\vect{P}^{*}$;\\
  14:~~~Given $\boldsymbol{\alpha}^{l+1}$, $\vect{B}^{l+1}$, and $\vect{P}^{l+1}$, transform the original problem (13) into problem (\ref{29});\\
  15:~~~Solve the problem (\ref{29}) and problem (\ref{33}) to obtain the optimal velocity of the UAV \\
  ~~~~~~~represented by $v^{l+1}$ and then the optimal trajectory $\vect Q^{l+1}$ can be obtained;\\
  16:~~~Update $l=l+1$;\\
  17:  \textbf{until}~~$\left|\eta(\boldsymbol {\alpha}^{l},\vect B^l,\vect P^l,\vect Q^l)-\eta(\boldsymbol {\alpha}^{l-1},\vect B^{l-1},\vect P^{l-1},\vect Q^{l-1})\right|$ $\leq \epsilon$ or $l\geq L_{\textmd{max}}$.\\
 % \textbf{Output:}\\
 % \arrayrulecolor{black}
  \hline
\end{tabular*}
\end{spacing}
\end{flushleft}
  \label{1}
\end{figure}

In step 15 of Algorithm 1, since $\mathcal{F}^{\textmd{lb},i}\left({X}_{k}[n]\right)$ is the lower bound of the first-order Taylor expansion of $\mathcal{F}\left({X}_{k}[n]\right)$ at the local point shown in (\ref{32}), the objective value of convex problem (\ref{33}) is a lower bound of problem (29). Thus, for given $\boldsymbol{\alpha}^{l+1}$, $\vect{B}^{l+1}$, $\vect{P}^{l+1}$, we have
\begin{equation}\label{36}
\Omega\left(\boldsymbol{\alpha}^{l+1}, \vect{B}^{l+1}, \vect{P}^{l+1}, \vect Q^{l}\right)\leq \Omega\left(\boldsymbol{\alpha}^{l+1}, \vect{B}^{l+1}, \vect{P}^{l+1}, \vect Q^{l+1}\right).
\end{equation}

The inequality (\ref{36}) indicates that although an approximate optimization problem (\ref{33}) is solved to obtain the UAV's trajectory $\vect Q$, the objective value of problem (29) is still non-decreasing after each iteration. Based on (\ref{34})-(\ref{36}), we obtain
\begin{equation}\label{37}
\Omega\left(\boldsymbol{\alpha}^{l}, \vect{B}^{l}, \vect{P}^{l}, \vect Q^{l}\right)\leq \Omega\left(\boldsymbol{\alpha}^{l+1}, \vect{B}^{l+1}, \vect{P}^{l+1}, \vect Q^{l+1}\right).
\end{equation}

The inequality (\ref{37}) suggests that the objective value of problem (\ref{13}) is non-decreasing after each iteration of Algorithm 1. Since the bandwidth and power in communication systems are limited, the objective value of problem (\ref{13}) is upper bounded by a finite value. Therefore, the proposed Algorithm 1 is guaranteed to converge.

In CVX, the interior-point method \cite{CVX} is usually invoked to solve optimization problems, and the calculation complexity of an algorithm mainly depends on the number of optimization variables and iterations. In this paper, the complexity of Algorithm 1 comes from three aspects. Firstly, in problem (\ref{15}), as the interior-point method is used to solve the relaxed user scheduling problem based on the given resource allocation and UAV trajectory, the computational complexity is $O\left(\log(1/\epsilon)(KN)^{3.5}\right)$, where $\epsilon$ is the given solution accuracy of Algorithm 1. Secondly, to solve problem (\ref{16}), the overall complexity is $O\left(M_{\textmd{inner}}M_{\textmd{outer}}\log(1/\epsilon)(KN)^{2}\right)$ according to the analytical expression in \cite{NOMA}, where $M_{\textmd{inner}}$ and $M_{\textmd{outer}}$ are the iteration numbers of the inner and outer loops of Lagrange duality. Finally, the complexity of solving problem (\ref{33}) with CVX is $O\left(\log(1/\epsilon)(KN)^{3.5}\right)$. Assuming $L$ is the  iteration number of the overall algorithm, the total complexity of Algorithm 1 can be calculated as
$O\left(L\log(1/\epsilon)\left(2(KN)^{3.5}+M_{\textmd{inner}}M_{\textmd{outer}}(KN)^{2}\right)\right)$.

\section{Simulation Results}\label{section:Simulation Results}

In this section, numerical results are provided to verify the effectiveness of joint user scheduling, resource allocation, and UAV trajectory control optimization algorithm in a UAV-enabled wireless network. We consider a system with 6 MGUs, which are randomly distributed on a horizontal plane. As explained in Section II, the origin of the 2D Cartesian coordinate is established at UC, and $\max\{\bar{r}_1/2,200\}$ is the initial flying radius of the UAV, where $\bar{r}_1$ is the distribution radius of MGUs in time slot $n=1$. Thus, the initial position of the UAV is set as $\vect{q}_{I}=(-\max\{\bar{r}_1/2,200\},0)$, while the calculation of UAV's final position $\vect{q}_{F}=(x_q,y_q)$ can be referred in Appendix A. We assume that the UAV flies at a fixed altitude $H=500$ m \cite{LiuJunyu,Wuqing}. The maximum transmission power of the UAV is $P_{\textmd{max}}=30$ dBm and the channel power gain at the reference distance $d_0=1$ m is $\rho_{0}=-50$ dB. The data rate threshold is set as $\gamma^{\textmd{th}}=8$ Mbps \cite{BeidaRelay} in a system with the available bandwidth $B_{\textmd{max}}=20$ MHz and the noise power spectrum density is $N_{0}=-169$ dBm/Hz. Furthermore, the minimum and maximum speeds of UAV are $V_{\textmd{min}}=20$ m/s and $V_{\textmd{max}}=100$ m/s, respectively. The Algorithm 1 convergence threshold is set as $\epsilon=10^{-3}$. If not specified otherwise, the flying trajectory of the UAV is sampled every time slot (i.e., $\delta=\frac{T}{N}=1$ s).

\begin{figure}[!t]
\centering
\includegraphics[width=9cm]{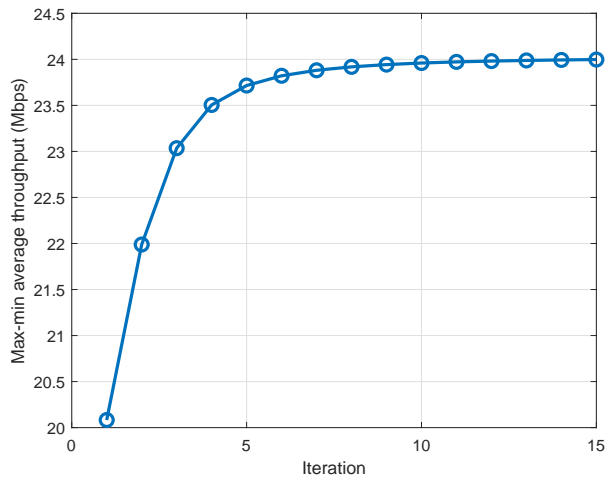}
\caption{The convergence of the proposed Algorithm 1 with period $T=120$ s.}
\label{Fig3}
\end{figure}

To ensure the feasibility of Algorithm 1, we first evaluate its convergence properties. In Fig. 2, it can be seen that at the beginning of the algorithm iteration, the max-min average throughput curve rises rapidly until the performance of the algorithm begins to converge. Besides, the proposed algorithm can reach convergence by just about 11 iterations, which indicates that Algorithm 1 can be effectively applied in practical applications.

\begin{figure}[!t]
\centering
\subfigure[The flying trajectories of the UAV within different periods $T$.]{
\label{Fig5.1}
\includegraphics[width=0.45\textwidth]{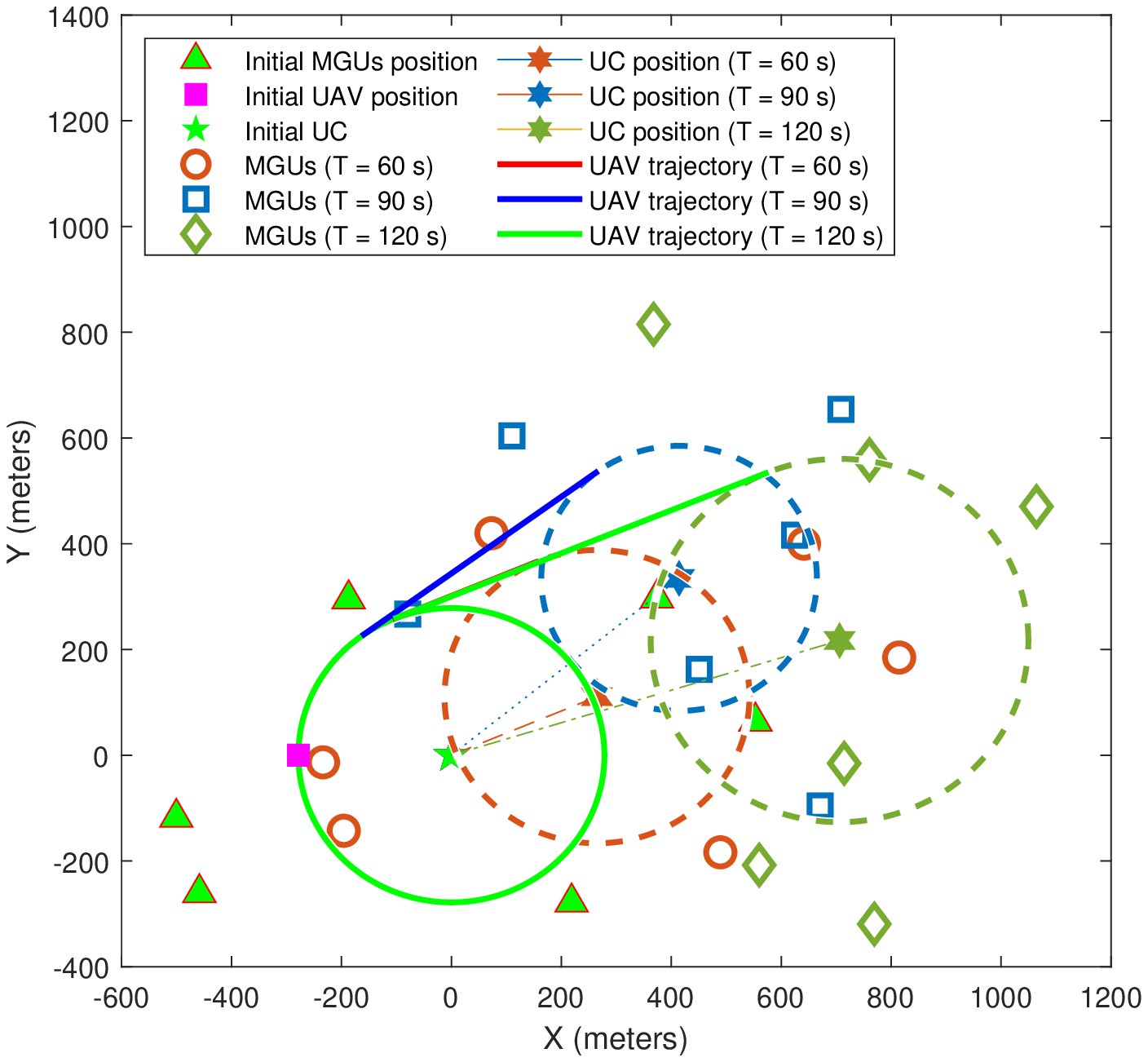}}
\subfigure[Period $T=60$ s.]{
\label{Fig5.2}
\includegraphics[width=0.45\textwidth]{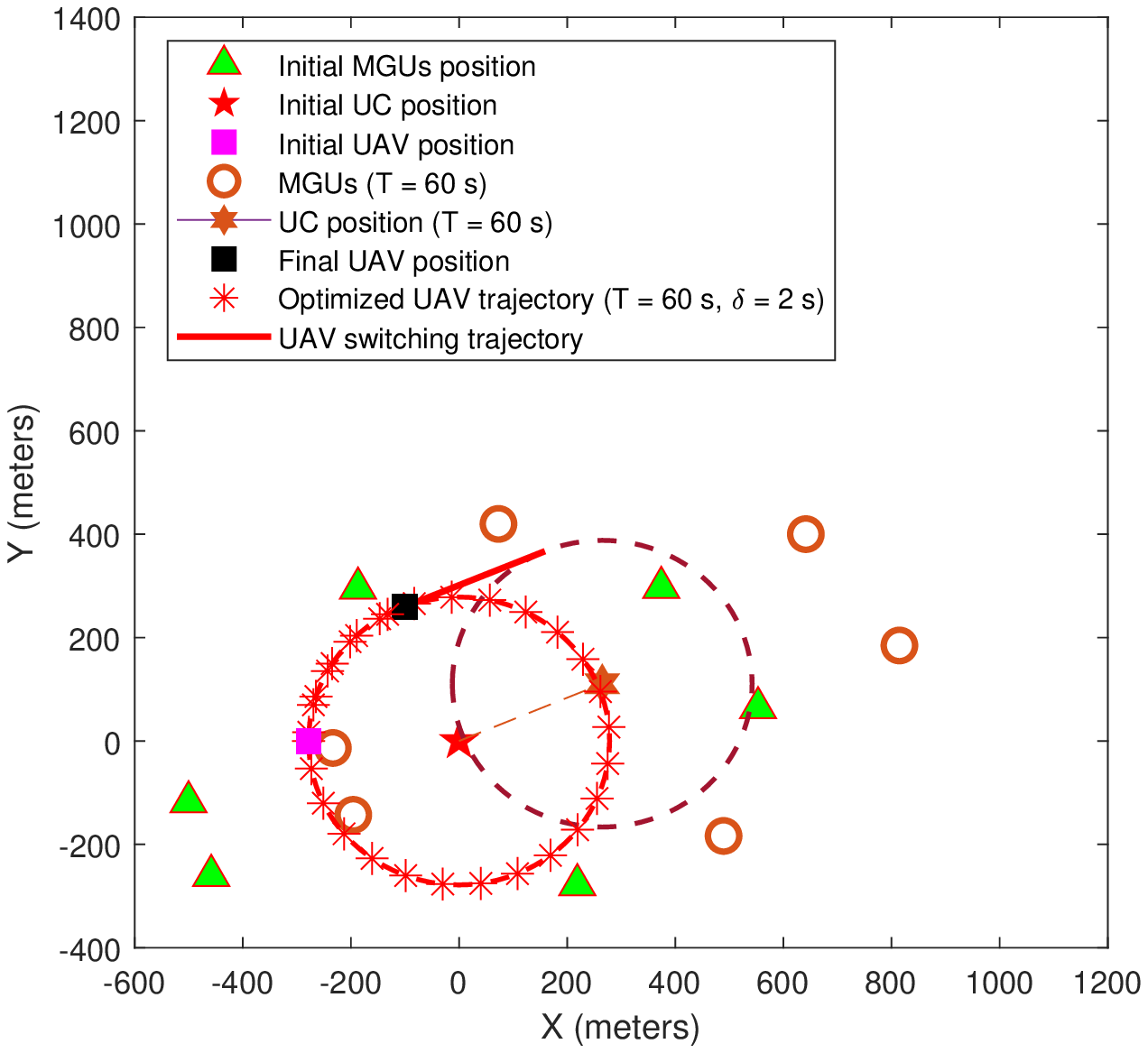}}
\subfigure[Period $T=120$ s.]{
\label{Fig5.3}
\includegraphics[width=0.45\textwidth]{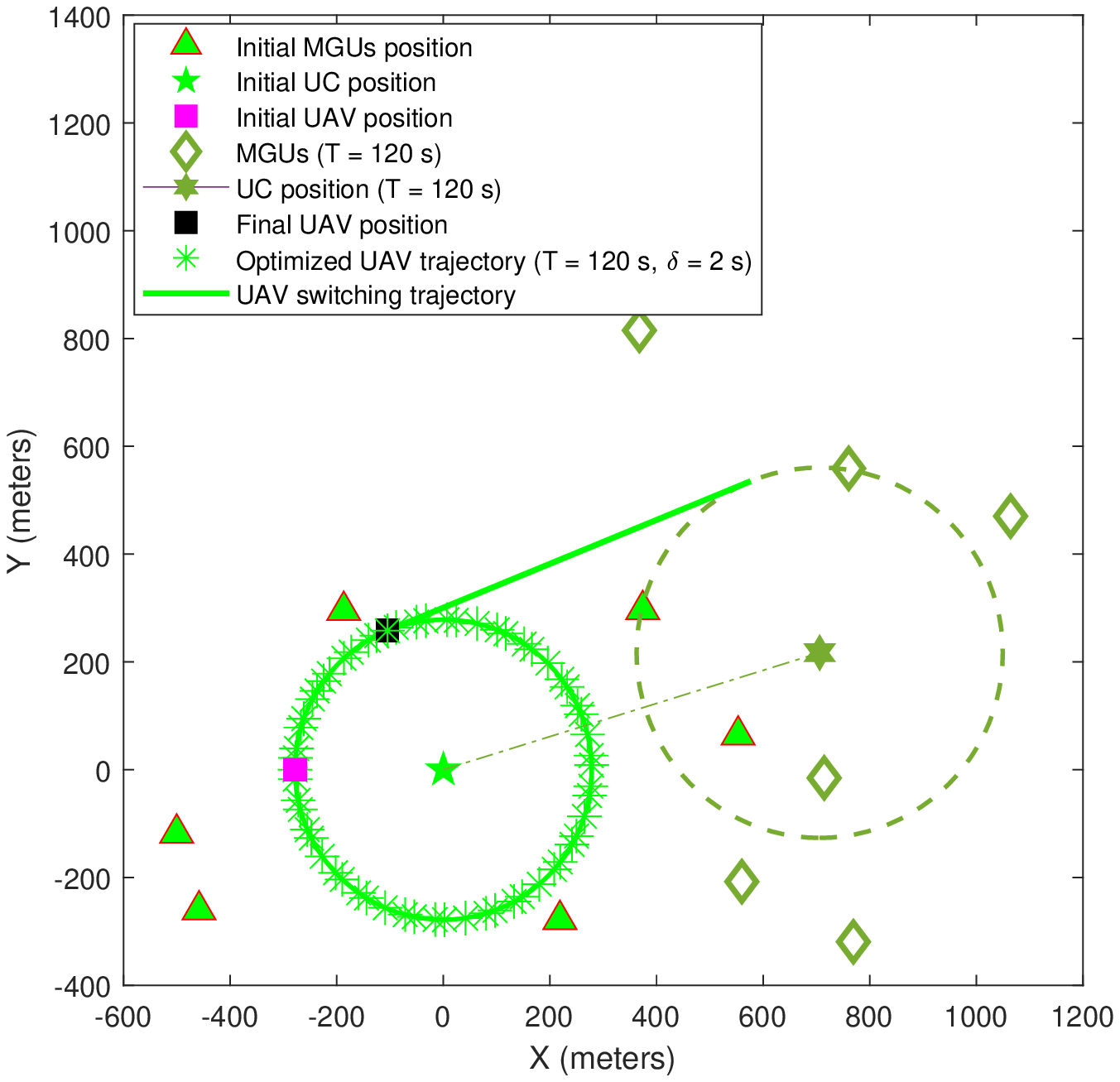}}
\caption{ Optimized trajectories of the UAV under different trajectory adjustment periods $T$ with time slot length $\delta=2$ s.}
\label{Fig5}
\end{figure}

In Fig. 3, we compare the optimal trajectories of the UAV under different trajectory adjustment periods $T$. All ground MGUs move in the area according to the RPGM model, while the UAV optimizes the trajectory based on our proposed Algorithm 1. In Fig. 3 (a), three optimized trajectories under different periods $T$ are depicted. Obviously, as the period $T$ changes, the moving distance and distribution of MGUs change accordingly. Therefore, the switching trajectories of the UAV in different periods $T$ are also different. Such change is the result of the optimization of Algorithm 1, which enables higher throughput of MGUs as much as possible. For clarity, the flying trajectory of the UAV is sampled every 2 s (i.e., $\delta=2$ s), and the sampled points are marked by '$\ast$'. In Fig. 3 (b), the UAV makes a detour from the initial position along the track of the red '$\ast$' mark to the switching point within the period $T=60$ s. During the period $T=120$ s shown in Fig. 3 (c), the UAV follows the green '$\ast$' track five times from the initial position to the switching point. Both of the optimization trajectories are obtained according to the users' motion to achieve the best LOS communication channel between the UAV and each MGU. In addition, the moving speed of ground MGUs is much lower than that of the fixed-wing UAV. When the period $T$ changes, the fixed-wing UAV optimizes the trajectory by adjusting the number of flight laps, i.e., by optimizing the UAV's velocity.

In Fig. 4, we analyze the relationship between the flight velocity $v$ and the number of laps $Cir$ of the UAV under different periods $T$. The position constraint of the switching trajectory point indicates that the number of flight laps $Cir$ must be an integer. Obviously, there is a positive correlation between the velocity $v$ and the number of laps $Cir$. The UAV's velocity increases by 29.15 m/s, 19.44 m/s, and 14.58 m/s as the number of flight laps adds one turn under periods $T=60$ s, $T=90$ s, and $T=120$ s, respectively. In this paper, it is worth noting that the minimum flight velocity $V_{\textmd{min}}=20$ m/s and the maximum flight velocity $V_{\textmd{max}}=100$ m/s of the fixed-wing UAV are constrained, reducing the velocity feasible domain. Furthermore, we can find that the longer the period $T$, the looser the feasible domain of the velocity $v$.

\begin{figure}[!t]
\centering
\includegraphics[width=9cm]{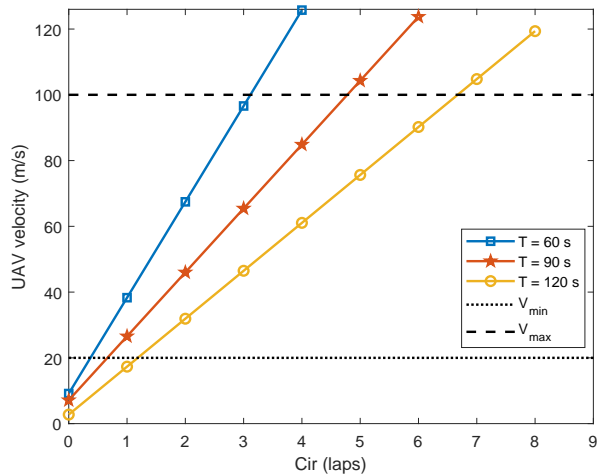}
\caption{ The optimized UAV velocity versus different laps $Cir$ under different periods $T$.}
\label{Fig6}
\end{figure}

\begin{figure}[!t]
\centering
\includegraphics[width=9cm]{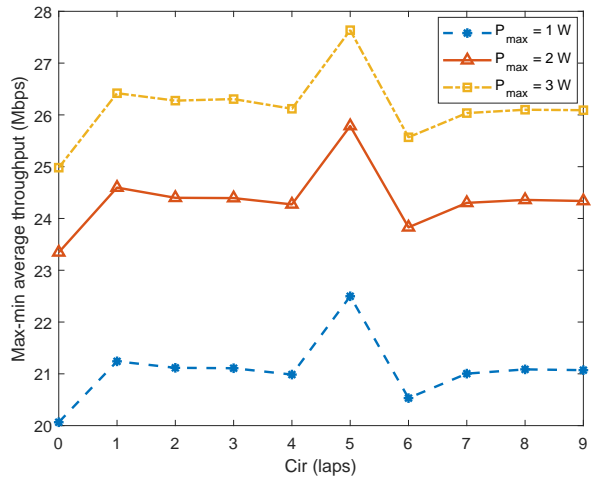}
\caption{ Max-min average throughput versus flying laps of the UAV with different maximum transmit power $P_{\textmd{max}}$.}
\label{Fig7}
\end{figure}

In Fig. 5, we investigate the influence of UAV flying laps and the maximum transmit power $P_{\textmd{max}}$ on system performance. It can be observed that the max-min average throughput is the highest when $Cir=5$. This is because the better channel gain can be obtained by optimizing the trajectory of the UAV in an air-ground system with dual mobility. Thus, opportunistic communication can be fully utilized to improve the throughput of MGUs. Moreover, the max-min average throughput is positively correlated with the maximum transmit power $P_{\textmd{max}}$ for two main reasons. On the one hand, with higher transmit power, the UAV can provide users with a stronger connection and a wider range of communication. On the other hand, increasing the transmission power means enriching the communication resources for users, which can increase the max-min average throughput as well.
\begin{figure}[!t]
\centering
\includegraphics[width=9cm]{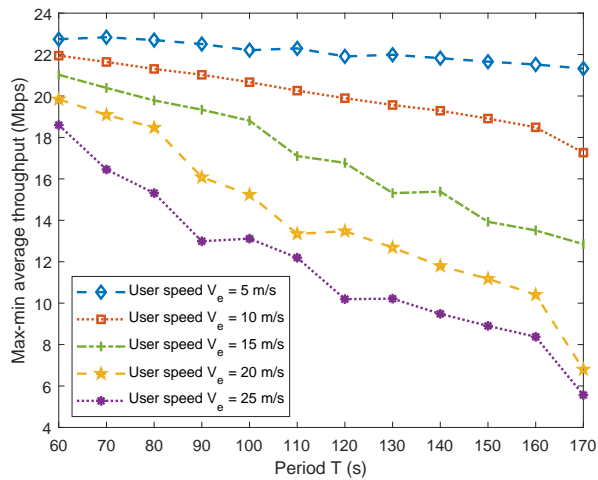}
\caption{  Max-min average throughput versus trajectory adjustment period $T$ with different users' speeds.}
\label{Fig8}
\end{figure}

In Fig. 6, we explore the impact of ground MGUs' speed and trajectory adjustment period $T$ on max-min average throughput. We observe an interesting phenomenon that as the period $T$ increases, the max-min average throughput of the system with the user speed of $V_e=5$ m/s fluctuates slightly. For instance, the performance of the system with period $T=70$ s is slightly higher than that of $T=60$ s, while the max-min average throughput corresponding to other speeds decreases all the time. This is because, with the slower speed of MGUs, the optimization of the UAV’s trajectory becomes more crucial to achieve better channel gain, especially in an air-ground system with dual mobility. However, as the user's moving speed increases, the influence of trajectory adjustment period $T$ is more obvious. In other words, when the period $T$ increases, the MGUs move farther and farther away from the initial UAV coverage circle, leading to a sharp decline in max-min average throughput performance. Thus, the UAV requires timely trajectory switching to improve the moving coverage performance for MGUs.
\begin{figure}[!t]
\centering
\includegraphics[width=9cm]{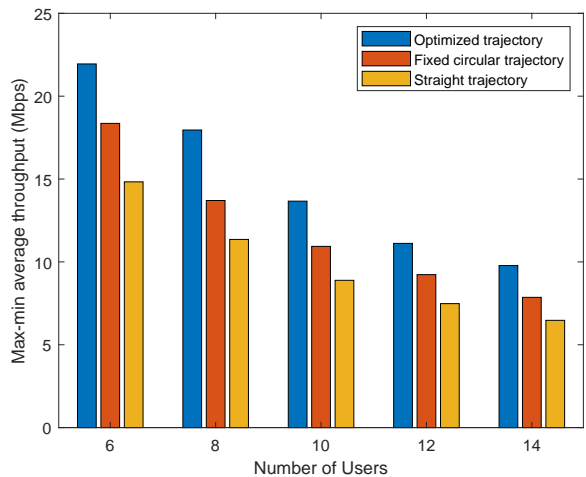}
\caption{Max-min average throughput versus the number of users with different trajectories.}
\label{Fig9}
\end{figure}

In Fig. 7, we compare the max-min average throughput achieved by the following three trajectories: 1) Optimized trajectory, which is obtained by Algorithm 1; 2) Fixed circular trajectory, where the UAV takes the center of the covered area as the flying center and flies in a circle with a radius of 600 m; and 3) Straight trajectory, where the UAV flies back and forth between the geometric center of the users' distribution locations at the initial time and the final time. Note that the round-trip time cost \cite{CurvatureLet} is included in the straight trajectory scheme, because the fixed-wing UAV cannot hover and turn around immediately. In addition, the user scheduling and resource allocation are optimized by Algorithm 1 with the given corresponding trajectory. As shown in Fig. 7, the optimized trajectory can achieve higher max-min average throughput than the other two trajectory schemes. One major reason is that the UAV can adjust the flight trajectory to obtain better channel gain. For these three trajectories, the max-min average throughput decreases with the increase of users. This is because the transmit power and bandwidth are assumed to be fixed in the system. As the number of users increases, the resources allocated to each user decrease, thereby reducing the max-min average throughput. Moreover, we can observe that the max-min average throughput gap between the three schemes is narrowing, as the number of users increases. Although the UAV can can improve the max-min average throughput by adjusting the flight trajectory, the effect of the trajectory control will become less and less obvious as the number of users increases.

In Fig. 8, we investigate the following three schemes: 1) Scheme I: All variables (i.e., user scheduling $\boldsymbol{\alpha}$, bandwidth allocation $\vect B$, power control $\vect P$, and UAV trajectory $\vect Q$) are jointly optimized by Algorithm 1; 2) Scheme II: The bandwidth allocation $\vect B$ and power control $\vect P$ are randomly optimized while the user scheduling and UAV trajectory are performed in the same way as in Algorithm 1; and 3) Scheme III: The variables $\boldsymbol{\alpha}$, $\vect B$, and $\vect P$ are randomly optimized while the UAV trajectory is optimized by Algorithm 1. Similar to Fig. 7, due to the limited communication resources, the max-min average throughput of these three schemes decreases with the increase of users. Compared with Scheme III, Scheme II optimizes the user scheduling according to the channel conditions between the UAV and MGUs, thus achieving higher throughput gains in the system with dual mobility. Compared with Scheme II, Scheme I shows obvious advantages, indicating that the resource optimization in Algorithm 1 is effective in improving the max-min average throughput performance of the communication system. In conclusion, Scheme I proposed in this paper has better adaptability to the multi-user system.

\begin{figure}[!t]
\centering
\includegraphics[width=9cm]{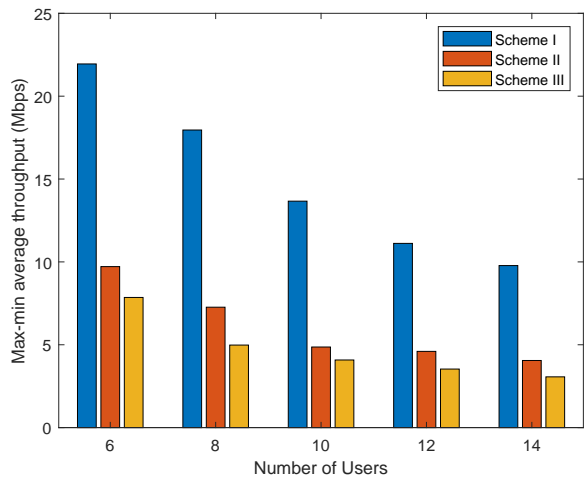}
\caption{Max-min average throughput versus the number of users with different optimization schemes.}
\label{Fig10}
\end{figure}

\section{Conclusion and future work}\label{section:conclusion}

In this paper, we have investigated a problem of applying the mobile fixed-wing UAV BS to provide moving coverage for MGUs. According to the flight characteristics of the fixed-wing UAV and the moving states of MGUs, a fixed-wing UAV-enabled wireless network architecture was proposed. Aiming at maximizing the minimum average throughput among users, we first applied variable relaxation, Lagrange dual, and SCA to optimize user scheduling, resource allocation, and UAV trajectory control, respectively. Subsequently, we proposed an efficient iterative algorithm to solve the challenging optimization problem. Extensive simulation results have shown that the proposed algorithm can provide excellent moving coverage performance in such UAV-enabled wireless network.

In future work, the optimization of flying radius and number of laps with the fixed flight velocity should be further explored to avoid sharp turnings that require large acceleration and deceleration. Since a single UAV may suffer from reliability problems, it is worth investigating the more general multi-UAV cooperative network based on intelligent mobile user clustering in future dynamic network planning. Finally, maximizing the energy-efficient of multi-UAV cooperative networks under the constraints of user mobility and communication requirements is also an interesting problem.

\appendices
\section{Derivation of the arc length $\wideparen{\vect{q}[n]\vect{q}[n+1]}$ and Dubins RSR trajectory switching tangent point $F$}

In Fig. 9, we assume that the center coordinates of circles $CR_{I}$ and $CR_{F}$ are $\left(x_{RI},y_{RI}\right)$ and $\left(x_{RF},y_{RF}\right)$, respectively. Thus, the equations of the two circles can be expressed as $\left(x-x_{RI}\right)^2+\left(y-y_{RI}\right)^2=r_{I}^2$ and $\left(x-x_{RF}\right)^2+\left(y-y_{RF}\right)^2=r_{F}^2$, where $r_{I}$ and $r_{F}$ denote the radius of $CR_{I}$ and $CR_{F}$, respectively. The angle $\hat{\theta}$ corresponding to the arc length of the UAV moving from time slot $n$ to time slot $n+1$, denoted as   $\wideparen{\vect{q}[n]\vect{q}[n+1]}$, can be obtained by the cosine theorem. Further, the arc length $\wideparen{\vect{q}[n]\vect{q}[n+1]}$ can be calculated as
\begin{equation}\label{arc}
\wideparen{\vect{q}[n]\vect{q}[n+1]}=r_I\cos^{-1}\left(\frac{2r_{I}^{2}-\left\|\vect{q}[n+1]-\vect{q}[n]\right\|^{2}}{2r_{I}^{2}}\right),~\forall n.
\end{equation}

Supposing the equation of the tangent line between these two circles is $y=Ax+B$. The distance between the center of two circles and the tangent line can be obtained as
\begin{equation}\label{38}
\left\{
\begin{array}{rcl}
\frac{\left|y_{RI}-Ax_{RI}-B\right|}{\sqrt{1+A^2}}=r_I,\\
\frac{\left|y_{RF}-Ax_{RF}-B\right|}{\sqrt{1+A^2}}=r_F.
\end{array}\right.
\end{equation}

\begin{figure}[!t]
\renewcommand{\figurename}{Fig.}
\centering
\includegraphics[width=9cm]{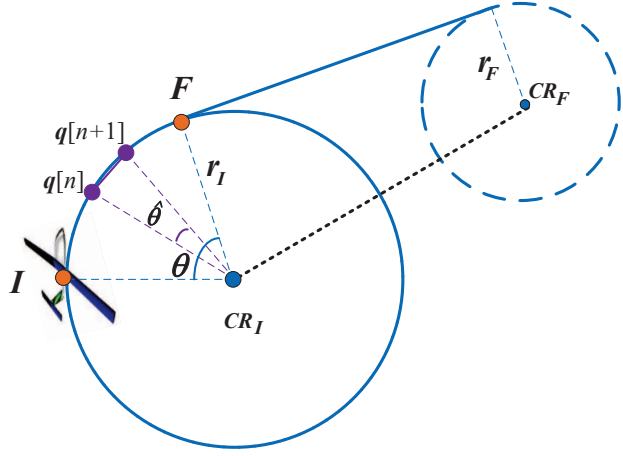}
\caption{A schematic illustration for the UAV trajectory.}
\label{Fig11}
\end{figure}

We assumed that the MGUs are heading towards or staying at the destination, but not in reverse motion. In other words, the UAV follows the Dubins RSR path so that the centers of the two circles fall to the lower right of the tangent, i.e.,
\begin{equation}\label{39}
y_{RI}-Ax_{RI}-B<0,~~ y_{RF}-Ax_{RF}-B<0.
\end{equation}

By substituting (\ref{39}) into (\ref{38}), we can obtain $A=\frac{-2CD\pm\sqrt{4C^2D^2-4\left(r_I^2-C^2\right)\left(r_I^2-D^2\right)}}{2\left(r_I^2-C^2\right)}$, where $C=x_{RI}-\frac{x_{RI}r_F-x_{RF}r_I}{r_F-r_I}$, $D=-y_{RI}-\frac{y_{RF}r_I-y_{RI}r_F}{r_F-r_I}$. However, when $A=\frac{-CD-\sqrt{C^2D^2-\left(r_I^2-C^2\right)\left(r_I^2-D^2\right)}}{\left(-r_I^2+C^2\right)}$, it corresponds to the Dubins right-straight-left (RSL) path \cite{DubinsICUAS}, which does not meet the requirements of this paper. Therefore, $A=\frac{-CD+\sqrt{C^2D^2-\left(r_I^2-C^2\right)\left(r_I^2-D^2\right)}}{\left(-r_I^2+C^2\right)}$. Then, we plug $A$ into the equation of the circle to obtain $B=\frac{\left(-y_{RF}+Ax_{RF}\right)r_I-\left(-y_{RI}+Ax_{RI}\right)r_F}{r_F-r_I}$. By combining the equations of the tangent and the circle, the coordinate of the tangent point $F=(x_q,y_q)$ can be calculated, where $x_q=\frac{x_{RI}-\left(B-y_{RI}\right)A}{A^2+1}$. Since the tangent point $F$ and the center of the circle $(x_{RI},y_{RI})$ are related to the equation shown as follows
\begin{equation}\label{40}
\left\{
\begin{array}{rcl}
x_q=x_{RI}-r_I\cos\theta,\\
y_q=y_{RI}+r_I\sin\theta,
\end{array}\right.
\end{equation}
thus $\theta=\cos^{-1}\left(\frac{x_{RI}-x_q}{r_I}\right)$ is determined. Substituting $\theta$ into (\ref{40}), we can get the coordinate of the tangent point $F$.

\section{Proof of Lemma 1}

From the definition of $\psi(x,y)$ in Lemma 1, we first obtain its second-order derivatives, which can be expressed as
\begin{equation}\label{41}
\psi_{xx}^{''}(x,y)=-\frac{1}{\ln 2}\frac{a^2xy^2}{x^2\left(x+ay\right)^2},
\end{equation}
\begin{equation}
\psi_{xy}^{''}(x,y)=\frac{ay}{\left(x+ay\right)\ln 2},
\end{equation}
\begin{equation}
\psi_{yx}^{''}(x,y)=\frac{a^2y}{\left(x+ay\right)^2\ln 2},
\end{equation}
\begin{equation}
\psi_{yy}^{''}(x,y)=-\frac{a^2x}{\left(x+ay\right)^2\ln 2}.
\end{equation}

Thus, the Hessian matrix of $\psi(x,y)$ is given by
\begin{equation}
 \boldsymbol{\psi}^{''}(x,y)=\left[ \begin{array}{cccc}
 -\frac{1}{\ln 2}\frac{a^2xy^2}{x^2\left(x+ay\right)^2} & \frac{ay}{\left(x+ay\right)\ln 2}\\
 \frac{a^2y}{\left(x+ay\right)^2\ln 2} & -\frac{a^2x}{\left(x+ay\right)^2\ln 2}
 \end{array} \right].
\end{equation}

For any matrix $\vect{t}=\left[t_1,t_2\right]^T$, where $[\cdot]^T$ represents the transpose operation, we have
\begin{equation}
\vect{t}^T\boldsymbol{\psi}^{''}(x,y)\vect{t}=-\frac{a^2}{x(x+ay)^2\ln 2}(yt_1-xt_2)^2.
\end{equation}

It is easy to observe that $\vect{t}^T\boldsymbol{\psi}^{''}(x,y)\vect{t}\leq 0$, when $a\geq0$, $x\geq0$ and $y\geq0$. Hence, the Hessian matrix $\boldsymbol{\psi}^{''}(x,y)$ is a negative semi-definite matrix and $\psi(x,y)$ is jointly concave with respect to $x$ and $y$.

\bibliography{IEEEabrv,reference}
\bibliographystyle{IEEEtran}

\end{document}